\documentclass[aps,pre,preprint]{revtex4-2}
\usepackage{amsmath,amssymb,amsfonts,theorem,color,bm}
\usepackage{graphics}
\usepackage{epsfig}

\begin{document}
\title{Hyperchaos, intermittency, noise and disorder in modified semiconductor superlattices}



\author{Luis L. Bonilla$^*$}
\affiliation{G. Mill\'an Institute for Fluid Dynamics, Nanoscience and Industrial Mathematics, and Department of Mathematics, Universidad Carlos III de Madrid, 28911 Legan\'es, Spain\\
$^{*}$Corresponding author: bonilla@ing.uc3m.es }

\author{ Manuel Carretero}
\affiliation{G. Mill\'an Institute for Fluid Dynamics, Nanoscience and Industrial Mathematics, and Department of Mathematics, Universidad Carlos III de Madrid, 28911 Legan\'es, Spain}

\author{Emanuel Momp\'o}
\affiliation{G. Mill\'an Institute for Fluid Dynamics, Nanoscience and Industrial Mathematics, and Department of Mathematics, Universidad Carlos III de Madrid, 28911 Legan\'es, Spain}




\affiliation{Departamento de Matem\'atica Aplicada, Universidad Pontificia Comillas, 28015 Madrid, Spain}





\begin{abstract}Weakly coupled semiconductor superlattices under dc voltage bias are nonlinear systems with many degrees of freedom whose nonlinearity is due to sequential tunneling of electrons. They may exhibit spontaneous chaos at room temperature and act as fast physical random number generator devices. Here we present a general sequential transport model with different voltage drops at quantum wells and barriers that includes noise and fluctuations due to the superlattice epitaxial growth. Excitability and oscillations of the current in superlattices with identical periods are due to nucleation and motion of charge dipole waves that form at the emitter contact when the current drops below a critical value. Insertion of wider wells increases superlattice excitability by allowing wave nucleation at the modified wells and more complex dynamics. Then hyperchaos and different types of intermittent chaos are possible on extended dc voltage ranges. Intrinsic shot and thermal noises and external noises produce minor effects on chaotic attractors. However, random disorder due to growth fluctuations may suppress any regular or chaotic current oscillations. Numerical simulations show that more than 70\% of samples remain chaotic when the standard deviation of their fluctuations due to epitaxial growth is below 0.024 nm (10\% of a single monolayer) whereas for 0.015 nm disorder suppresses chaos.\end{abstract}

\maketitle

\section{Introduction}\label{intro} 
Semiconductor superlattices (SSLs) are artificial crystals made out of a regular periodic array of layers of two different semiconductors stacked on top of each other \cite{esa70,esa74,all86,gra95,bon05}. These semiconductors have different band gaps and similar lattice constants, so that the conduction band edge of an infinitely long ideal SSL is a succession of quantum wells (QWs) and quantum barriers (QBs). When the latter are wide enough, QWs are weakly coupled and the resulting SSL behaves as an excitable or oscillatory system depending on its configuration \cite{bon05}. Excitability and self-sustained oscillations involve the generation of charge dipole waves inside the SSL and their motion \cite{bon05}. Devices made out of SSLs include oscillators \cite{sch97,min98,pal99,xu10}, detectors \cite{win98,win00,kla01,sha18}, quantum cascade lasers \cite{fai94,wac02prb,mal05,ter07,vit15,bel15,bon05}, and all-electronic fast generators of true random numbers \cite{li13,mom21}. The latter are crucial to secure fast and safe data storage and transmission \cite{sti95,gal08,nie00}, stochastic modeling \cite{asm07}, and Monte Carlo simulations \cite{bin02}. The origin of randomness for true random number generators based on SSLs is quantum partition noise due to electron tunneling \cite{bla00,rey03,fev18}. However, the unpredictability of the final number sequence is due to chaotic evolution (described by deterministic model equations), which amplifies a random quantum seed into a fast process that achieves generation rates up to hundreds of Gb/s. SSL devices are smaller and more scalable than similarly fast optoelectronic devices based on random semiconductor lasers \cite{uch08,rei09,kan10,sci15,tan15}. In these devices, chaotic processes in the laser amplify quantum noise, which is the origin of randomness. Most of the unpredictability of the final random number sequence rests on chaotic evolution, which is deterministic. In a sense, random SSLs and semiconductor lasers work as physical pseudorandom number generators that take a random quantum seed and expand these small fluctuations at the quantum level into a fast changing physical process that achieves generation rates up to hundreds of Gb/s.

While early experiments observed chaos in GaAs/AlAs SSLs at ultralow temperatures \cite{zha96,gra99}, spontaneous current oscillations, quasiperiodic attractors and chaos have been observed at room temperature in GaAs/Al$_{0.45}$Ga$_{0.55}$As SSLs \cite{hua12}. 
External noise may induce or enhance chaotic oscillations over a wider voltage bias range provided its amplitude is sufficiently large and its bandwidth is much smaller than the oscillation frequency \cite{yin17}. Appropriate external noise of sufficient strength can induce self-sustained oscillations in an otherwise stationary state (coherence resonance \cite{pik97,gia00,mar04,lee05,hiz06}) and help detecting a superimposed weak periodic signal by stochastic resonance \cite{mom18,sha18}; see also Refs.~\cite{fau83,mcn88,gam98,han02,bad05,bur08} for stochastic resonance.

To achieve a better understanding and control of SSL based random number generators, we need to improve our theoretical explanations of spontaneous chaos at room temperature. SSLs are nonlinear systems with many degrees of freedom, whose effective nonlinearity originates from the well-to-well sequential resonant tunneling process \cite{sch89,gra90,gra91,bon02r,bon05}. Most sequential tunneling models of electron transport consider {\em ideal} SSLs with identical periods \cite{bon02r,bon05,wac02,BT10}. Numerical simulations of ideal SSLs have shown that spontaneous chaos exists on narrow intervals of voltage bias and it is enhanced by noise \cite{alv14,bon17,yin17}. Also, short SSLs  at room temperature display clear period doubling cascades to chaos, which occur on shorter voltage intervals for longer SSLs \cite{ess18}. Period doubling routes to spontaneous chaos at ultralow temperatures were predicted earlier \cite{ama02,ama03}. Random imperfections strongly affect spontaneous chaos in SSLs \cite{ess18}. Overall, spontaneous chaos predicted by numerical simulations of ideal SSLs exists on shorter voltage interval than reported in experiments. 

More recently, we have put forward the idea that a systematic modification of SSL design produces more robust spontaneous chaos at room temperature \cite{mom21}. The idea is to {\em design appropriate imperfections} in SSLs by inserting two identical and wider QWs in the SSL. In ideal SSLs, self-sustained current oscillations are due to repeated generation of dipole waves at the emitter contact and motion through the SSL to the collector contact. We show that inserting one wider QW may trigger dipole waves in it. In turn, complex dynamics arises out of the competition of two identical wider QWs as nucleation sites of dipole waves. We find hyperchaos, chaos with more than one positive Lyapunov exponent, and intermittent chaos due to random triggering of dipole waves. The connection of Lyapunov exponents to the Kolmogorov-Sinai metric entropy can be found in Refs.~\cite{ott93,cen10}; see also Ref.~\cite{yuk99} for its generalization and use in statistical analysis of time series. We also study the effect of imperfections and noise on this design of chaotic SSLs and show that it is robust.  

The rest of the paper is organized as follows. We describe the deterministic version of our microscopic sequential model of ideal SSL electron transport in Section \ref{sec:2} and Appendix \ref{app1}. Since we want our model to be realistic, we use a detailed model with different effective masses and voltage drops at wells and barriers \cite{gol87b,gol87a,agu97}. Section \ref{sec:3} discusses the $I-V$ current-voltage characteristic curve of an ideal SSL based on this model. For appropriate values of the emitter contact conductivity and other parameters, self-sustained oscillations of the current appear for a certain interval of bias voltages. In Section \ref{sec:4}, we study the changes on the model equations due to having imperfect barriers and wells with varying widths and to internal and external noise. In Section \ref{sec:5}, we include a single wider well or two wider wells in a SSL and study the resulting changes on the SSL current-voltage characteristics by numerical simulations of the deterministic equations. Details of hyperchaos and intermittent chaos are given in Section \ref{sec:6}. For designing chaotic SSLs, it is important to study the influence of noise and randomness in the obtained enhanced spontaneous chaos. This is done in Section \ref{sec:7}. The last section contains the summary and conclusions of this work. 

\section{Microscopic sequential tunneling model}
\label{sec:2}
Commonly used models of electron transport in SSLs based on sequential tunneling are reviewed in Refs.~\cite{bon02r,wac02,bon05,BT10,bon17}. Experimental confirmation in weakly coupled SSLs is abundantly documented in Ref.~\cite{bon05}. In sequential resonant tunneling models, each SSL period is described by average values of the electric field and electron density. The effective masses and permittivities of the different materials comprising the SSL are replaced by average values. The resulting models are discrete in space. The importance of using spatially discrete equations was recognized in early models, which were motivated by the formation of stationary electric field domains in SSLs \cite{lik89,lai91,lai93,pre94,bon94}. In this paper, we treat the barriers and wells as separate entities \cite{agu97,bon00}, seeking a less symmetric representation of the SSL that may give a more complete and realistic description of electron transport and spontaneous chaos. The resulting model is more complete than those considered in these previous works \cite{agu97,bon00}. 

\subsection{Rate equations for subband populations}
The main charge transport mechanism in a weakly coupled SL is sequential resonant tunneling. We assume that the intrasubband scattering time is much shorter than the intersubband scattering time which, in turn, is much shorter than the interwell tunneling time across barriers. Typically, the time scale for carrier thermalization within a subband is 0.1 ps, the carriers reach thermal equilibrium with the lattice in times smaller than 100 ps, the tunneling time is about 500 ps and the time scale associated with self-sustained oscillations of the current is longer than 10 ns \cite{bon94}. In processes varying on the latter scale, there is a local equilibrium Fermi-Dirac distribution at each subband $\nu=1,\ldots, n$ at the lattice temperature (for the numerical parameters used in numerical simulations, $n=3$) with 2D electron densities $n_i^{(\nu)}$ related to their chemical potentials $\mu_i^{(\nu)}$ by \cite{bon02r}
\begin{eqnarray}
n_i^{(\nu)} = \frac{m_Wk_BT}{\pi\hbar^2}\int_0^\infty A_{C\nu}(\epsilon)\, \ln\!\left(1+e^{(\mu_i^{(\nu)}-\epsilon)/k_BT}\right) d\epsilon.   \label{eq1}
\end{eqnarray}
Here $i=1,\ldots, N$, where $N$ is the number of SL periods. For the time being, we assume that all SL periods $d_B+d_W$ ($d_B$ and $d_W$ are the widths of barrier and wells, respectively) are identical (otherwise we have to label the widths of barriers and wells with indices) and that the electron temperature at each subband $\nu$ of energy $\mathcal{E}_{C\nu}$ (measured from the bottom of the $i$th well) equals the lattice temperature $T$. $m_W$ and $k_B$ are the electron effective mass at the wells and the Boltzmann constant, respectively. Scattering is included in our model by means of Lorentzian functions: 
\begin{eqnarray}
A_{C\nu}(\epsilon) = \frac{\gamma_\nu}{\pi}\,\frac{1}{(\epsilon -
\mathcal{E}_{C\nu})^{2} +\gamma_\nu^{2}} \label{eq2}
\end{eqnarray}
(for the $i$th well). The Lorentzian half-width is $\gamma_\nu=\hbar/\tau_{sc}$, where $\tau_{sc}$ is the lifetime associated to any scattering process dominant in the sample (interface roughness, impurity scattering, phonon scattering\ldots) \cite{wac97b,bon00}. Of course this phenomenological treatment of scattering could be improved by calculating microscopically the self-energy associated to one of the scattering processes mentioned above \cite{wac99,wac97a}, or even exchange-correlation effects (which affect the electron charge distribution in a self-consistent way). However restricting ourselves to a single scattering mechanism would result in a loss of generality and simplicity of the model.
The electronic states of a SSL with infinite lateral extension have wave functions $e^{iqx+\mathbf{k}_\perp\cdot\mathbf{x}_\perp}u_q(x)$ (a plane wave on the lateral directions $\mathbf{x}_\perp=(y,z)$ times a Bloch state on the direction of the superlattice vertical growth; $u_q(x)$ is a periodic function of $x$ with the SSL period). The energy minibands $\epsilon(q)$ corresponding to the previous Bloch states solve a 1D Kronig-Penney model \cite{bastard,wei91}
\begin{subequations}\label{eq3}
\begin{eqnarray}
&&\cos ql= \cos kd_W \cosh\alpha d_B-\frac{1}{2}\!\left(\frac{1}{\xi}-\xi\right)\!\sin kd_W \sinh\alpha d_B, \label{eq3a}\\ 
&& k =\frac{\sqrt{2 m_{W}\epsilon}}{\hbar},\quad\alpha = \frac{\sqrt{2 m_B (eV_{B} - \epsilon)}}{\hbar},\quad \xi=\frac{m_W\alpha}{m_B k}=\sqrt{\frac{m_W}{m_B}\!\left(\frac{eV_B}{\epsilon}-1\right)}. \label{eq3b}
\end{eqnarray}
\end{subequations}
In the limit as $\alpha d_B\to\infty$, Eq.~\eqref{eq3a} produces the subbands $\epsilon=\mathcal{E}_{C\nu}$ appearing in Eq.~\eqref{eq2}:
\begin{eqnarray}
\cos kd_W -\frac{1}{2}\!\left(\frac{1}{\xi}-\xi\right)\!\sin kd_W =0, \label{eq4}
\end{eqnarray}
where $k$ is given by Eq.~\eqref{eq3b}. Using the symmetry of the quantum well \cite{bastard}, Eq.~\eqref{eq4} factorizes into two equations corresponding to even and odd wave functions, respectively: 
\begin{eqnarray*}
\cos\frac{kd_W}{2} -\frac{1}{\xi}\sin\frac{kd_W}{2}=0,\quad \cos\frac{kd_W}{2}+\xi\sin\frac{kd_W}{2} =0.
\end{eqnarray*}

We shall write the rate equations for the electron densities and $n=3$ to simplify the number of tunneling channels:
\begin{subequations}\label{eq5}
\begin{eqnarray}
\dot{n}_i^{(1)} &=& \frac{1}{e}[J_{1,i-1\to 1,i}-J_{1,i\to 1,i+1}-J_{1,i\to 2,i+1}-\ldots-J_{1,i\to n,i+1}]+\frac{n_i^{(2)} }{\tau_{21}}+\ldots+\frac{n_i^{(n)} }{\tau_{n1}}, \quad \label{eq5a}\\
\dot{n}_i^{(2)} &=&\frac{1}{e}[J_{1,i-1\to 2,i}-J_{2,i\to 3,i+1}]-\frac{n_i^{(2)} }{\tau_{21}}, \label{eq5b}\\
\dot{n}_i^{(3)} &=&\frac{1}{e}[J_{1,i-1\to n,i}+J_{2,i-1\to 3,i+1}]-\frac{n_i^{(3)} }{\tau_{31}},  \label{eq5c}
\end{eqnarray}
\end{subequations}
where $\tau_{21}$, $\tau_{31}$ are the intersubband scattering times within a quantum well, $-e<0$ is the charge of the electron, and $J_{\nu,i\to \nu',i+1}$ is the current density from subband $\nu$ of QW $i$ to subband $\nu'$ of QW $i+1$. The current densities may be approximately calculated by means of the Transfer Hamiltonian method\cite{bardeen,pay86,bre88,jon89,pla89,ina96,pla04}. It is important to note that the current densities are functions of the local electric field that exhibit several peaks when all the electron densities equal the doping density \cite{bon02r,bon05}. See details in Appendix \ref{app1}; here we only quote the results  \cite{agu97,bon02r}.

\subsection{Tunneling current }
Let $J_{1,i\to\nu, i+1}$ be the tunneling current through the $i$th barrier from the first subband of QW $i$ to the $\nu$th subband of QW $i+1$. As explained in Appendix \ref{app1}, we have \cite{bon02r,bon17}
\begin{eqnarray}
&&J_{1,i\to\nu,i+1} = \frac{e\hbar k_{B}T}{2m_{B}}
\int_0^\infty A_{C1}(\epsilon)\, A_{C\nu}\!\left(\epsilon+eV_i+\frac{e}{2}(V_{w_i}+V_{w_{i+1}})\right)
B_{i-1,i}(\epsilon)  \,\,\nonumber\\
&&\quad\quad\quad\quad\times B_{i,i+1}(\epsilon)\, T_{i}(\epsilon)\, \ln\!\left[\frac{1+e^{\frac{\mu_{i}^{(1)}-\epsilon}{k_{B}T}}}{1+e^{\frac{\mu_{i+1}^{(\nu)}-eV_i-e(V_{w_i}+V_{w_{i+1}})/2-\epsilon}{k_{B}T}}} \right] d\epsilon, \label{eq6}
\end{eqnarray}
in which the energies $\epsilon$ are measured from the bottom of the $i$th well, and:
\begin{itemize}
\item $B_{i-1,i}$ are given by
\begin{subequations}\label{eq7}
\begin{eqnarray}
&&B_{i-1,i} = k_{i}\,\left[d_W + \left(\frac{1}{\alpha_{i-1}} +
\frac{1}{\alpha_{i}}\right)\!
\left(\frac{m_B}{ m_{W}} \sin^{2} \frac{k_{i}d_W}{ 2} + \cos^{2}\frac{k_{i}d_W}{2}\right)\!\right]^{-1}\!,
\label{eq7a}\\
&&\hbar k_{i} =\sqrt{2 m_{W}\epsilon},\quad \hbar k_{i+1} = \sqrt{2 m_{W} e\left(\frac{\epsilon}{e}+ V_{i} + \frac{V_{w_{i}}+V_{w_{i+1}}}{2}\right)},\label{eq7b}\\
&&\hbar\alpha_{i} = \sqrt{2 m_B e\!\left(V_{B} - \frac{V_{w_{i}}}{ 2} - \frac{\epsilon}{e} \right)},\quad
\hbar\alpha_{i-1} = \sqrt{2 m_B e \left(V_{B} + \frac{V_{w_{i}}}{2} + V_{i-1} - \frac{\epsilon}{e}\right)},\quad \label{eq7c}\\
&&\hbar\alpha_{i+1} = \sqrt{2 m_B e\left(V_{B}  - \frac{V_{w_{i}}}{2} - V_{i} - V_{w_{i+1}} -
\frac{\epsilon}{e}\right)},  \label{eq7d} 
\end{eqnarray}
\end{subequations}
where $k_{i}$ and $\alpha_{i}$ are the wave numbers in the wells and the barriers, respectively, $m_{W}$ and $m_B$ are the effective masses of the electrons at the wells and barriers, respectively, and $d_W$ and $d_B$ are the widths of wells and barriers, respectively. $\hbar B_{i,i+1}/m_B$ are the attempt frequencies related to sequential tunneling through the $i$th barrier. $V_i$ and $V_{w_{i}}$, $i=1,\ldots,N$, are the potential drops at the $i$th barrier and well, respectively. We assume that the potential drops at barrier and wells are non-negative and that the electrons are singularly concentrated on a plane located at the end of each well. Then $k_i$ (dictated by the Transfer Hamiltonian method, cf Appendix \ref{app1}) depends on the electric potential at the center of the $i$th well, whereas $\alpha_{i}$ depends on the potential at the beginning of the $i$th barrier, $V_{w_{i}}/2$. The potential drops $V_0$ and $V_N$ correspond to the barriers separating the SL from the emitter and collector contacts, respectively. $e V_{B}$ is the barrier height in the absence of potential drops.  
\item $T_i$ is the dimensionless transmission probability through the $i$th barrier separating wells $i$ and $i+1$: 
\begin{eqnarray}
T_{i}(\epsilon) = \frac{1}{\frac{(k_{i}+k_{i+1})^2}{4k_{i}k_{i+1}} +\frac{1}{4}\left(\frac{m_Bk_{i}}{m_W\alpha_i} +\frac{m_W\alpha_{i}}{m_Bk_i}\right)\!
\left(\frac{m_B k_{i+1}}{m_W\alpha_{i}} + \frac{m_{W}\alpha_{i}}{m_Bk_{i+1}}\right)\!\sinh^2(\alpha_id_B)}\,. \label{eq8}
\end{eqnarray}
\end{itemize}

\subsection{Poisson equations}
The voltage drops through the structure are calculated as
follows. The Poisson equation yields the potential drops in the
barriers, $V_{i}$, and the wells, $V_{wi}$:
\begin{subequations}\label{eq9}
\begin{eqnarray}
\varepsilon_{W}\,\frac{V_{w_{i}}}{d_W}
&=&\varepsilon_B\,\frac{V_{i-1}}{d_B}+\frac{e}{2}\left(n_i-N_{D}\right)\!,\quad n_i=\sum_{\nu=1}^nn_{i}^{(\nu)},
\label{eq9a}\\
\varepsilon_B\frac{V_{i}}{d_B}&=&\varepsilon_B\frac{V_{i-1}}{d_B}+e\left(n_i-N_{D}\right)\!,\quad  i=1,\ldots,N, \label{eq9b}
\end{eqnarray}
\end{subequations}
where $\varepsilon_W$ and $\varepsilon_B$ and $N_D$ are the well and barrier static permittivities and the 2D intentional doping density at the wells, respectively \cite{gol87b,gol87a,agu97,bon00}.

\subsection{Boundary conditions}
In Ref.~\cite{bon00}, the boundary conditions consist of using the current densities of Eqs.~\eqref{eq6}, \eqref{eq7} and \eqref{eq8} in an Amp\`ere's law derived from Eqs.~\eqref{eq5}, together with a model of the emitter and collector layers and a simplified version of the metal semiconductor contact. In this work, we shall use simpler phenomenological Ohm laws for the current density at emitter and collector, namely
\begin{subequations}\label{eq10}
\begin{eqnarray}
&&J_{0\to 1}= \sigma_e\frac{V_0}{d_{B_0}}, \label{eq10a}\\
&&J_{N\to N+1}=\sigma_c\frac{n_N}{N_{D_N}}\,\frac{V_N}{d_{B_N}}. \label{eq10b}
\end{eqnarray}
\end{subequations}
Here $\sigma_j$, $j=e,c$ are the contact conductivities, $d_{B_j}$ are effective lengths for the contact regions and $N_{D_N}$ is an effective 2D doping density of the collector, cf. Ref.~\cite{bon02r}.

The condition of overall voltage bias between contacts closes the set of equations:
\begin{eqnarray}
V_\text{dc} &=& \sum_{i=0}^{N}V_{i}+\sum_{i=1}^{N}V_{wi}. \label{eq11}
\end{eqnarray}

\subsection{Elimination of the potential drops at the wells}
The previous model gives rise to many equations but some of them are not independent. We can eliminate the potential drops at the wells from the system, as done in Ref.~\cite{bon00}. Eqs.~(\ref{eq9}) imply
\begin{eqnarray}
\frac{\varepsilon_{W}V_{w_{i}}}{ \varepsilon_B d_W} = \frac{V_{i-1} + V_{i}}{ 2d_B}, \quad i=1,\ldots,N.\label{eq12}
\end{eqnarray}
Then the bias condition (\ref{eq11}) becomes
\begin{eqnarray}
V_\text{dc} &=&\!\left(1+\frac{\varepsilon_B d_W}{ \varepsilon_{W}d_B}\right)\!\sum_{i=0}^{N}V_{i} - \frac{\varepsilon_B (V_{0}+V_{N})\, d_W}{2\varepsilon_{W} d_B}. \label{eq13}
\end{eqnarray}
Instead of the rate equations (\ref{eq5}), we can derive a form of Amp\`ere's law which explicitly contains the total current density $J(t)$. We differentiate Eq.~(\ref{eq9b}) with respect to time and eliminate $n_i=\sum_{\nu=1}^3n_i^{(\nu)}$ by using Eqs.~(\ref{eq5}). The result is
\begin{eqnarray}
&& \frac{\varepsilon_B}{d_B}\frac{dV_{i}}{dt} + J_{i\to i+1} = J(t),\label{eq14}\\
&& J_{i\to i+1}=\sum_{\nu=1}^3J_{1,i\to \nu, i+1}+ J_{2,i\to 3, i+1}, \label{eq15}
\end{eqnarray}
where $i=0,1,\ldots,N$ and the total current density $J(t)$ is the sum of displacement and tunneling currents. If there are more than 3 subbands, Eq.~\eqref{eq15} will include more contributions from other tunneling channels.

\subsection{Elimination of the higher subband populations}
Typically $\tau_{21}$ and $\tau_{31}$ in Eqs.~(\ref{eq5}) are much smaller than the dielectric relaxation time. Then Eqs.~(\ref{eq5b}) and (\ref{eq5c}) have the quasi-stationary solutions
\begin{eqnarray}
n_i^{(2)}\approx\frac{\tau_{21}}{e}J_{1,i-1\to 2,i}, \ldots, \quad
n_i^{(n)} \approx\frac{\tau_{n1}}{e}J_{1,i-1\to n,i},  \label{eq16}
\end{eqnarray}
and $n_i^{(2)},\ldots, n_i^{(n)}\ll n_i^{(1)}$. Then we can set $n_i^{(2)}=\ldots=n_i^{(n)}=0$, $n_i^{(1)}=n_i$, $\mu_{i}^{(1)}=\mu_{i}$, given by Eq.~(\ref{eq1}), 
\begin{eqnarray}
n_i = \frac{m_Wk_BT}{\pi\hbar^2}\int_0^\infty A_{C1}(\epsilon)\, \ln\!\left(1+e^{(\mu_i-\epsilon)/k_BT}\right) d\epsilon,   \label{eq17}
\end{eqnarray}
$\mu_i^{(\nu)}\to-\infty$ for $\nu\neq 1$, and Eq.~(\ref{eq15}) becomes 
\begin{eqnarray}
&&J_{i\to i+1}=J_{i\to i+1}^+(V_{i-1},V_i,V_{i+1},\mu_{i},T)- J_{i\to i+1}^-(V_{i-1},V_i,V_{i+1},\mu_{i+1},T), \label{eq18}
\end{eqnarray}
where
\begin{subequations}\label{eq19}
\begin{eqnarray}
&&J_{i\to i+1}^+=\frac{e\hbar k_{B}T}{2m_{B}}\sum_{\nu=1}^n \int_0^\infty A_{C1}(\epsilon)\, A_{C\nu}\!\left(\epsilon+eV_i+\frac{ed_W\varepsilon_B}{4d_B\varepsilon_W}(V_{i-1}+V_{i+1}+2V_{i})\right) \nonumber \\ 
&&\quad\quad\quad\,\times\, B_{i-1,i}(\epsilon)\, B_{i,i+1}(\epsilon)\, T_{i}(\epsilon)\, \ln\!\left(1+e^{\frac{\mu_{i}-\epsilon}{k_{B}T}}\right) d\epsilon, \label{eq19a}\\
&& J_{i\to i+1}^-=\frac{e\hbar k_{B}T}{2m_{B}}\int_0^\infty\! A_{C1}(\epsilon)\, A_{C1}\!\left(\epsilon+eV_i+\frac{ed_W\varepsilon_B}{4d_B\varepsilon_W}(V_{i-1}+V_{i+1}+2V_{i})\!\right)\! B_{i-1,i}(\epsilon) \nonumber \\ 
&&\,\times B_{i,i+1}(\epsilon) T_{i}(\epsilon)\ln\!\left[1+ \exp\!\left(\frac{1}{k_{B}T}\!\left(\mu_{i+1} -\epsilon-eV_i-ed_W\varepsilon_B\frac{V_{i-1}+V_{i+1}+2V_{i}}{4d_B\varepsilon_W}\right)\!\!\right)\!\right]\!\!d\epsilon. \,\quad\quad\label{eq19b}
\end{eqnarray}
\end{subequations}
Since all tunneling currents from subbands with $\nu>1$ are negligible, Eqs.~\eqref{eq19} hold for any number of subbands, not only for $n=3$. These equations differ from the usual sequential tunneling model that includes a sum over higher subbands in Eq.~(\ref{eq19b}). 

The time-dependent model consists of the $3N+2$ equations (\ref{eq9b}), (\ref{eq13}), (\ref{eq14}), (\ref{eq17}) [the currents are given by Eqs.~(\ref{eq18}) and (\ref{eq19})],
which contain the $3N+2$ unknowns $n_{j}$, $\mu_j$ ($j=1,\ldots,N$),
$V_j$ ($j=0,1,\ldots,N$), and $J$. Thus we have a system of equations which, together with appropriate initial conditions, determine completely and self-consistently the voltage drops, current density, and electron densities. For convenience, let us list again the minimal set of equations we need to solve in order to determine completely all the unknowns:
\begin{subequations}\label{eq20}
\begin{eqnarray}
&&\frac{\varepsilon_B}{d_B}\frac{dV_{i}}{dt} + J_{i\to i+1} = J(t),
\quad\quad\quad\quad i=0,1,\ldots,N , \label{eq20a}\\
&&\varepsilon_B\frac{V_{i}}{d_B} = \varepsilon_B\frac{V_{i-1}}{d_B}+ e\, (n_{i}
- N_{D}),\quad i=1,\ldots,N , \label{eq20b}\\
&& n_i = \frac{m_Wk_BT}{\pi\hbar^2}\int_0^\infty A_{C1}(\epsilon)\, \ln\!\left(1+e^{(\mu_i-\epsilon)/k_BT}\right) d\epsilon, \quad i=1,\ldots,N,  \label{eq20c}\\
&&V_\text{dc} = \left(1+\frac{\varepsilon_Bd_W}{\varepsilon_{W}d_B}\right)
\sum_{i=0}^{N}V_{i} - \frac{\varepsilon_Bd_W}{2\varepsilon_{W}d_B}(V_0+V_{N}), \label{eq20d}
\end{eqnarray}\end{subequations}
together with the constitutive relations given by Eqs.~(\ref{eq10}), \eqref{eq18} and \eqref{eq19}. 

\section{Current-voltage characteristics and attractors for ideal SSL}\label{sec:3}
In this section, we review the different stable configurations that may appear in ideal SSLs with identical periods described by Eqs.~\eqref{eq20} and \eqref{eq18}-\eqref{eq19}. 

\begin{figure}[htbp]
\begin{center}
\includegraphics[width=12cm,angle=0]{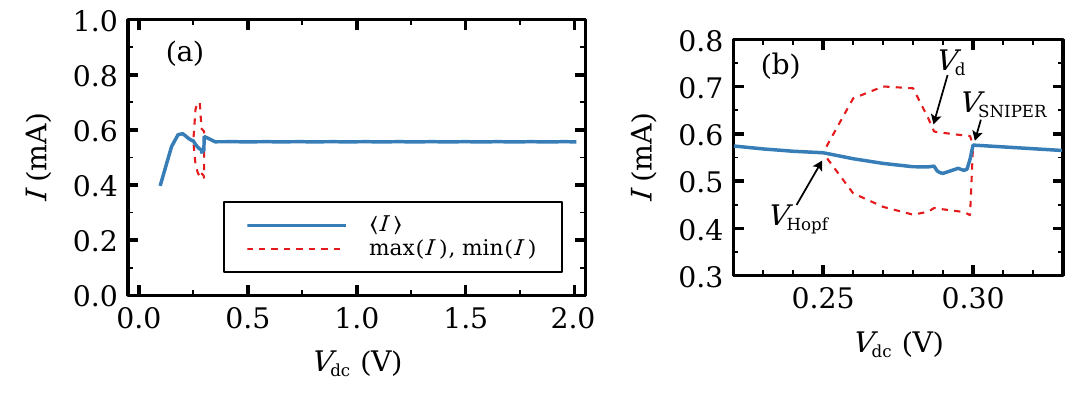}\\
\includegraphics[width=7cm,angle=0]{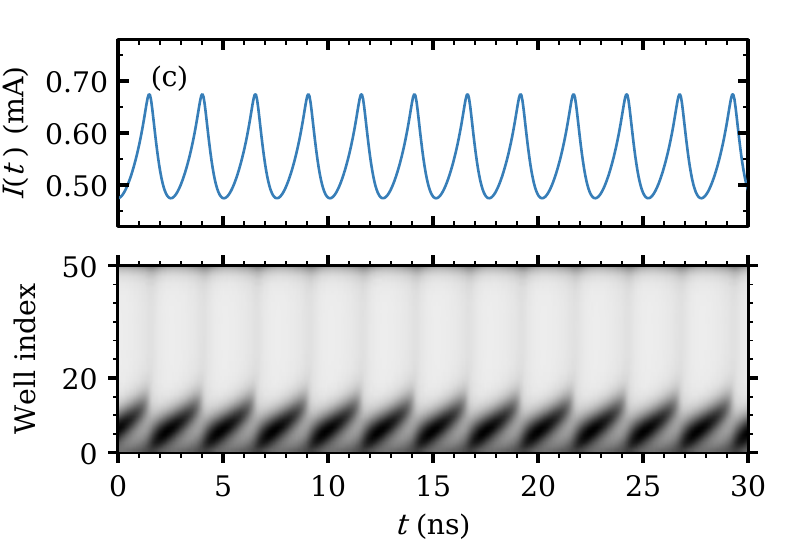}~~~
\includegraphics[width=7cm,angle=0]{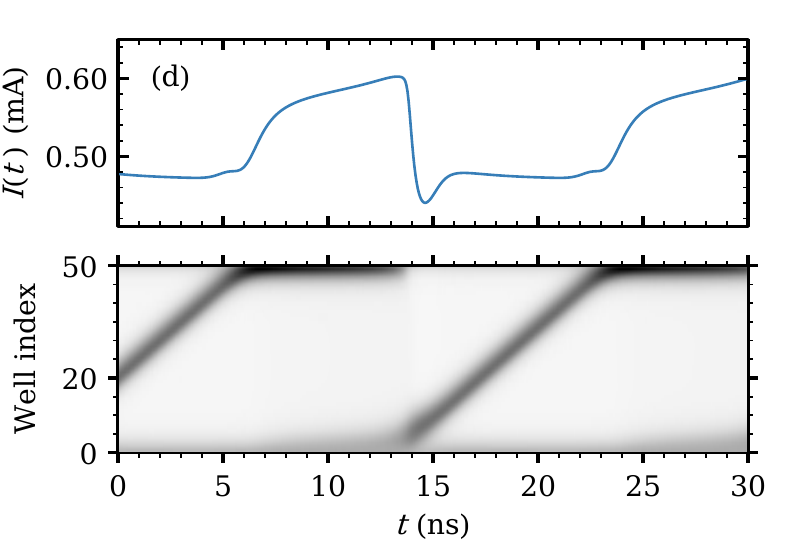}
\end{center}
\caption{(\textbf{a}) First plateau of the current-voltage characteristics for the SSL with 50 identical periods. (\textbf{b}) Zoom of the region of self-oscillations that appear as a supercritical Hopf bifurcation and end at a SNIPER. $V_d$ is the voltage at which the frequency of the oscillations drop to lower values. At $V_d$, the high electric field domains move throughout the SSL and reach the collector contact. Current traces and the corresponding density plots of the electric field for (\textbf{c}) \(V_{dc}=0.26\,\text{V}\) (just after the Hopf bifurcation) and (\textbf{d}) \(V_{dc}=0.29\,\text{V}\) (just before the SNIPER bifurcation). Light and dark tones correspond to low and high field values, respectively.} 
\label{fig1} 
\end{figure}

Our basic SSL configuration is that of References \cite{li13,hua12}: An ideal SSL with $N=50$ periods whose three relevant subband energies, 41.6, 165.8, and 354.3 meV, are calculated by means of Eq.~\eqref{eq8}. The level broadenings due to scattering are 2.5, 8 and 24 meV, respectively, for the three subbands involved in the $I-V$ characteristics we study \cite{mom21}. The equivalent 2D doping density due to the doping of the central part of the quantum well is $N_D=6 \times 10^{10}$ cm$^{-2}$. Also $m_W=0.063\, m_e$, $m_B=(0.063 + 0.083x)m_e=0.1 m_e$ (for $x=0.45$), $A=s^2$ with $s=30\,\mu$m, $d_B=4$ nm, $d_W=7$ nm, $l=d_B+d_W$, $\varepsilon_B= 10.9 \epsilon_0$, $\varepsilon_W=(12.9-2.84x) \epsilon_0$, $\epsilon_0$, and $V_\text{dc}$ are the effective electron mass at wells and barriers, the SL cross section, the side length of a square mesa, the (Al,Ga)As barrier thickness, the GaAs well thickness, the SL period, the barrier permittivity, the well permittivity, the dielectric constant of the vacuum, and the dc voltage bias, respectively. We select contact conductivities $\sigma_c=\sigma_e= 0.49$ A/Vm and the same doping density $N_D$ for injector and collector. 

The current-voltage $I-V$ curve of the SSL gives an overall picture of the different stable configurations thereof as a function of the applied {\em dc} voltage $V$, which acts as a control parameter. The $I-V$ curve has regions of increasing current separated by almost plateaus  between them. The regions of increasing current roughly correspond to subband energies of Eq.~\eqref{eq8} counted from the first one, $(\mathcal{E}_{C\nu}-\mathcal{E}_{C1})/e$, and the plateaus to the intervals between them. Fig.~\ref{fig1} shows the first plateau. For this ideal SSL, the current increases linearly from zero voltage (not shown) until it reaches the first plateau. The stable solution of Eqs.~\eqref{eq20} with boundary conditions \eqref{eq10} is time independent except for a voltage region of time-periodic solutions whose maxima and minima and average current are marked in Fig.~\ref{fig1}(a) and \ref{fig1}(b). The stationary solutions at the plateau are frozen wavefronts that increase with QW index from a low to a high value of the voltage drops $V_i$ (which equal the local electric field). In the limit of an infinitely long SSL, $J_{i\to i+1}=J$ for $n_i=N_D$ and $V_i=F$ has three solutions $F^{1)}<F^{(2)}<F^{(3)}$, and $F^{(1)}$ and $F^{(3)}$ are the low and high field values, respectively. Depending on the value of the stationary current density $J$, a wavefront on the infinitely long SSL does not move (it is pinned by the lattice) or it moves with constant velocity \cite{bon02r,BT10}. For a SSL with finitely many periods, $J$ is fixed by the bias condition Eq.~\eqref{eq20d} \cite{bon02r,bon05}.

\begin{figure}[htbp]
\begin{center}
\includegraphics[width=12cm,angle=0]{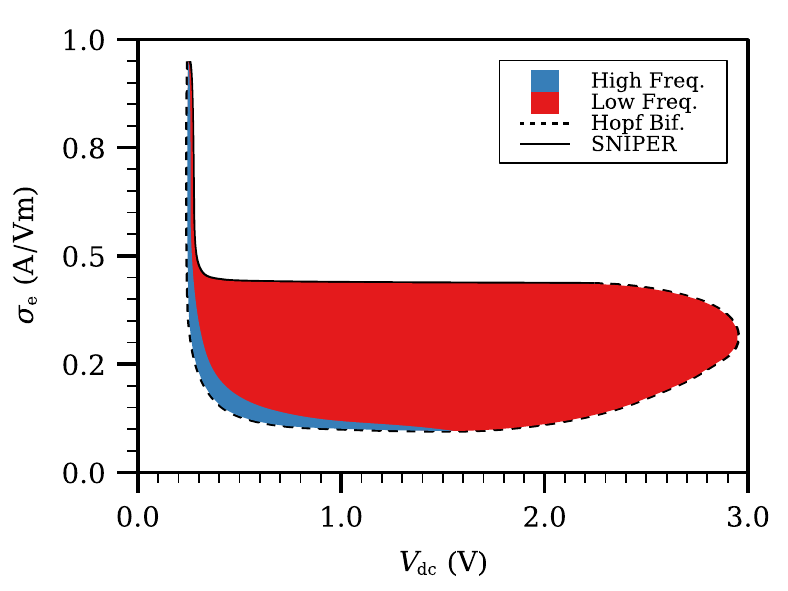}
\end{center}
\caption{Phase diagram of injector contact conductivity versus dc voltage exhibiting a bounded region of current self-oscillations. At the dashed boundary line, the self-oscillations appear as Hopf bifurcations from the stationary field profile which is linearly stable outside the bounded region. The continuous boundary line corresponds to oscillations disappearing at a saddle-node infinite period bifurcation, as selected in the main text. In the red regions, self-oscillations have low frequency and correspond to fully formed charge dipole waves that move across the entire SSL. In the blue regions, high frequency self-oscillations correspond to charge dipole waves that disappear before reaching the receiving contact. Reprinted from E. Momp\'o, M. Carretero, L. L. Bonilla, Designing hyperchaos and intermittency in semiconductor superlattices, Physical Review Letters 127, 096601 (2021); https://doi.org/10.1103/PhysRevLett.127.096601} \label{fig2} 
\end{figure}

The branch of time-periodic solutions exists provided the number of SSL periods is 14 or higher for the parameters of our SSL, cf Refs.~\cite{bon94,wac97,car00} for theory on a simpler model. The branch of time-periodic solutions starts as a supercritical Hopf bifurcation and ends at a saddle-node infinite period (SNIPER) bifurcation, i.e., it ends at finite amplitude and infinite period by collision of the periodic attractor with a homoclinic orbit. The time-periodic self-sustained oscillations are caused by the repeated formation of dipole waves (traveling high-field domains) at the emitter, motion toward and annihilation at the collector, as seen in Figs.~\ref{fig1}(c) and \ref{fig1}(d) for voltages near the beginning (Hopf) and end (SNIPER) of their voltage region. Fig.~\ref{fig1}(c) and \ref{fig1}(d) show current traces and the corresponding density plot of the electric field inside the SSL. Near the Hopf bifurcation, the high-field domains are repeatedly born at the emitter contact and die before reaching the collector contact as seen in Fig.~\ref{fig1}(c), which produces high-frequency oscillations. For $V_\text{dc}>V_d$, the domains reach the collector as seen in Fig.~\ref{fig1}(d), and the oscillation frequency drops. The frequency vanishes at the SNIPER bifurcation.

Fig.~\ref{fig2} contains the phase diagram of the emitter contact $\sigma_e$ of Eq.~\eqref{eq10}(a) as a function of $V_\text{dc}$. We observe that self-sustained oscillations are possible in a region of the parameter plane $(V_\text{dc},\sigma_e)$. Below a certain $\sigma_e$, self-sustained oscillations begin and end with Hopf bifurcations, whereas for larger contact conductivity, they end at a SNIPER bifurcation, cf Fig.~\ref{fig1}. We have adopted a value in the latter region because it produces results compatible with experiments.

\section{Noise, imperfections and superlattice configurations}\label{sec:4}
\subsection{Stochastic equations}
Internal and external noises can be included in our model as indicated in Ref.~\cite{bon17}. We add noise terms to Eqs.~\eqref{eq20a} and \eqref{eq20d}, 
\begin{eqnarray}
&&\frac{\varepsilon_B}{d_B}\frac{dV_{i}}{dt}+ J_{i\to i+1}+\xi_i(t)=J(t),\quad i=0,\ldots, N,\label{eq21}\\
&&V_\text{dc} = \left(1+\frac{\varepsilon_Bd_W}{\varepsilon_{W}d_B}\right)
\sum_{i=0}^{N}V_{i} - \frac{\varepsilon_Bd_W}{
2\varepsilon_{W}d_B}(V_0+V_{N})+\eta(t), \label{eq22}
\end{eqnarray}
keeping the other equations unchanged and including a voltage fluctuation $\eta(t)$ \cite{mom18}. The fluctuations of the current density are independent identically distributed (i.i.d.) zero-mean white noises with correlations:
\begin{eqnarray}
\langle\xi_i(t)\xi_j(t')\rangle&=&\frac{e}{A}\!\left[J_{i\to i+1}^+(V_{i-1},V_i,V_{i+1},\mu_{i},T)+ J_{i\to i+1}^-(V_{i-1},V_i,V_{i+1},\mu_{i+1},T)\right.\nonumber\\
&+&\left.2k_BT\frac{\partial}{\partial\mu_i}J_{i\to i+1}^-(V_{i-1},V_i,V_{i+1},\mu_{i},T) \right]\!\delta_{ij}\delta(t-t'). \label{eq23}
\end{eqnarray}
We have assumed that the internal noise is due to shot and thermal noise \cite{bla00}. The first two terms in the right hand side of Eq.~(\ref{eq23}) are due to shot noise \cite{bla00} and the last one to thermal fluctuations. We model current fluctuations associated to dissipation due to electron diffusion by Landau-Lifshitz fluctuating hydrodynamics \cite{ll6,kei87} adapted to SLs. Eq.~(\ref{eq18}) can be written as a discrete drift-diffusion current density,
\begin{eqnarray}
J_{i\to i+1}&=&J_{i\to i+1}^+(V_{i-1},V_i,V_{i+1},\mu_{i},T)-J_{i\to i+1}^-(V_{i-1},V_i,V_{i+1},\mu_{i},T)\nonumber\\
&-&[ J_{i\to i+1}^-(V_{i-1},V_i,V_{i+1},\mu_{i+1},T)- J_{i\to i+1}^-(V_{i-1},V_i,V_{i+1},\mu_{i},T)],\label{eq24}
\end{eqnarray}
where the last two terms correspond to electron diffusion \cite{bon17}. Considered as a a function of the chemical potential and fixing the voltage drops, this discrete diffusion yields
$$\delta[J_{i\to i+1}^-(\mu_{i+1})-J_{i\to i+1}^-(\mu_{i})]\approx\frac{\partial J_{i\to i+1}^-(\mu_i)}{\partial\mu_i} \,\delta(\mu_{i+1}-\mu_i).$$
Then the fluctuations of the current density are i.i.d.\ zero-mean white noises with correlation given by the last term in Eq.~(\ref{eq23}), cf. Ref.~\cite{kei87}. This is similar to fluctuations in Gunn diodes where the diffusion current is proportional to the diffusion coefficient times the electron density \cite{kei87}, instead of the nonlinear expression in Eq.~(\ref{eq23}).

\subsection{Non-ideal superlattices}
Typically, there are doping density fluctuations at the wells, one-monolayer fluctuations of barrier and well widths and fluctuations in Al concentration at barriers. In addition, we may want to change the width of some wells and explore how the SSL dynamics changes. While it is clear that doping density fluctuations affect Poisson equations, the other fluctuations modify importantly the tunneling currents. Local changes in $d_B$, $d_W$, $m_B$, $\varepsilon_B$ and $V_B$ change the location of energy levels at each well, the barrier wave numbers $\alpha_i$, and the coefficient functions $B_i$, $T_i$ and $A_{Cj}$. Fluctuations in doping density, barrier and well width affect electrostatics and Amp\`ere's law. 

\subsubsection{Electrostatics and Amp\`ere's law}
Equations (\ref{eq9a}) and (\ref{eq9b}) become
\begin{subequations}\label{eq25}
\begin{eqnarray}
\varepsilon_{W}\,\frac{V_{w_{i}}}{d_{W_i}}
&=&\varepsilon_{B_{i-1}}\,\frac{V_{i-1}}{d_{B_{i-1}}}+\frac{e}{2}\left(n_i-N_{Di}\right)\!,
\label{eq25a}\\
\varepsilon_{B_i}\frac{V_{i}}{d_{B_i}}&=&\varepsilon_{B_{i-1}}\frac{V_{i-1}}{d_{B_{i-1}}}+e\left(n_i-N_{Di}\right)\!, \label{eq25b}
\end{eqnarray}\end{subequations}
from which we obtain,
\begin{eqnarray}
V_{w_{i}}= \frac{d_{W_i}}{2\varepsilon_W}\!\left(\varepsilon_{B_{i-1}}\,\frac{V_{i-1}}{d_{B_{i-1}}}+\varepsilon_{B_i}\frac{V_{i}}{d_{B_i}}\right)\!. \label{eq26}
\end{eqnarray}
The electron density of Eq.~(\ref{eq20d}) becomes
\begin{subequations} \label{eq27}
\begin{eqnarray}
&&n_i = \frac{m_Wk_BT}{\pi\hbar^2}\int_0^\infty A_{C1_i}(\epsilon)\, \ln\!\left(1+e^{(\mu_i-\epsilon)/k_BT}\right) d\epsilon,\label{eq27a}\\
&&A_{C\nu_i}(\epsilon) = \frac{\gamma_\nu}{\pi}\,\frac{1}{(\epsilon - \mathcal{E}_{C\nu_i})^{2} +\gamma_\nu^{2}}.\label{eq27b} 
\end{eqnarray}\end{subequations}
Amp\`ere's law can be obtained in the same way as Eq.~(\ref{eq14}). Now it is:
\begin{eqnarray}
 \frac{\varepsilon_{B_i}}{d_{B_i}}\frac{dV_{i}}{dt} + J_{i\to i+1} +\xi_i(t) = J(t). \label{eq28}
 \end{eqnarray}
The i.i.d. zero-mean white noises $\xi_i(t)$ have correlations given by Eq.~(\ref{eq23}). The voltage bias condition of Eq.~(\ref{eq22}) is now
 \begin{eqnarray}
V_\text{dc}\! =\!\sum_{i=0}^{N}\!\left(1+\frac{d_{W_i}+d_{W_{i+1}}}{2 \varepsilon_{W}d_{B_i}}\varepsilon_{B_i} \right)\! V_{i} \!-\! \frac{\varepsilon_{B_0}d_{W_0}V_{0}}{2\varepsilon_{W} d_{B_0}} \!-\! \frac{\varepsilon_{B_N}d_{W_{N+1}}V_{N}}{2\varepsilon_{W} d_{B_N}}\! +\!\eta(t).  \label{eq29}
\end{eqnarray}
The total current $J(t)$ can be calculated from Eq.~(\ref{eq28}) and the bias condition in Eq.~(\ref{eq29}), thereby providing effective nonlocal equations of motion when substituted back in Eqs.~\eqref{eq28}. 
 
\subsubsection{Tunneling currents}
The disorder modifies the energy levels measured from the bottom of each well so that they depend on the well number and we denote them as $\mathcal{E}_{Cj_i}$. Barrier effective masses, permittivities, widths and wave numbers are also modified. The tunneling current densities given by Eqs.~(\ref{eq19}) become
\begin{subequations}\label{eq30}
\begin{eqnarray}
&&J_{i\to i+1}^+\!=\!\frac{e\hbar k_{B}T}{2m_{B_i}}\! \sum_{\nu=1}^3\!
\int_0^\infty \! A_{C1_i}(\epsilon) B_{i-1,i}(\epsilon)\, B_{i,i+1}(\epsilon)\, T_{i}(\epsilon)\, \ln\!\left(1+e^{\frac{\mu_{i}-\epsilon}{k_{B}T}}\right) \nonumber \\ 
&&\quad\times\, A_{C\nu_{i+1}}\!\!\left(\epsilon+\!\left(\!1+\frac{\varepsilon_{B_i}}{d_{B_i}}\frac{d_{W_i}+d_{W_{i+1}}}{4\varepsilon_W}\!\right)\!eV_i\!+\frac{d_{W_i}\varepsilon_{B_{i-1}}eV_{i-1}}{4d_{B_{i-1}}\varepsilon_W}+\frac{d_{W_{i+1}}\varepsilon_{B_{i+1}}eV_{i+1}}{4d_{B_{i+1}}\varepsilon_W}\!\right)\!  d\epsilon,\quad\quad\label{eq30a}\\
&& J_{i\to i+1}^-=\frac{e\hbar k_{B}T}{2m_{B_i}}\int_0^\infty\! A_{C1_i}(\epsilon)\, B_{i-1,i}(\epsilon)\, B_{i,i+1}(\epsilon)\, T_{i}(\epsilon) \nonumber\\
&&\quad\times\, A_{C1_{i+1}}\!\left(\epsilon+\!\left(1+\frac{\varepsilon_{B_i}}{d_{B_i}}\frac{d_{W_i}+d_{W_{i+1}}}{4\varepsilon_W}\right)\!eV_i\!+\frac{d_{W_i}\varepsilon_{B_{i-1}}eV_{i-1}}{4d_{B_{i-1}}\varepsilon_W}+\frac{d_{W_{i+1}}\varepsilon_{B_{i+1}}eV_{i+1}}{4d_{B_{i+1}}\varepsilon_W}\!\right)\!  \nonumber \\ 
&&\quad\times 
\ln\!\left[1+ \exp\!\left(\frac{\mu_{i+1} -\epsilon}{k_BT}-\!\left(1+\frac{\varepsilon_{B_i}}{d_{B_i}}\frac{d_{W_i}+d_{W_{i+1}}}{4\varepsilon_W}\right)\frac{eV_i}{k_BT}\!-\frac{d_{W_i}\varepsilon_{B_{i-1}}eV_{i-1}}{4k_BTd_{B_{i-1}}\varepsilon_W}\right.\right.\nonumber\\
&&\quad\,\left.\left.-\frac{d_{W_{i+1}}\varepsilon_{B_{i+1}}eV_{i+1}}{4k_BTd_{B_{i+1}}\varepsilon_W}\!\right)\!\right] d\epsilon,\label{eq30b}
\end{eqnarray}
\end{subequations}
for $i=1,\ldots,N-1$. 

\section{$I-V$ characteristic curves of modified superlattices} \label{sec:5}
In this section, we ignore noise and fluctuations in doping density and in barrier and well widths. We discuss how introducing one or two wider wells changes the current-voltage characteristics $I-V$ of an otherwise ideal SSL. Fig.~\ref{fig1}(a) and \ref{fig1}(b) show the $I-V$ curve of the ideal SSL with appropriate emitter contact conductivity. It exhibits self-sustained oscillations of the current in a narrow voltage region provided the number of SSL periods is 14 or larger. These oscillations are periodic in time and are caused by the formation of traveling regions of high field (which are charge dipole waves) at the emitter, motion toward and annihilation at the collector; see Fig.~\ref{fig1}(c) and \ref{fig1}(d). For the chosen value of $\sigma_e$, the branch of oscillations starts as a supercritical Hopf bifurcation and ends at a SNIPER bifurcation.  

\begin{figure}[htbp]
\begin{center}
\includegraphics[width=12cm,angle=0]{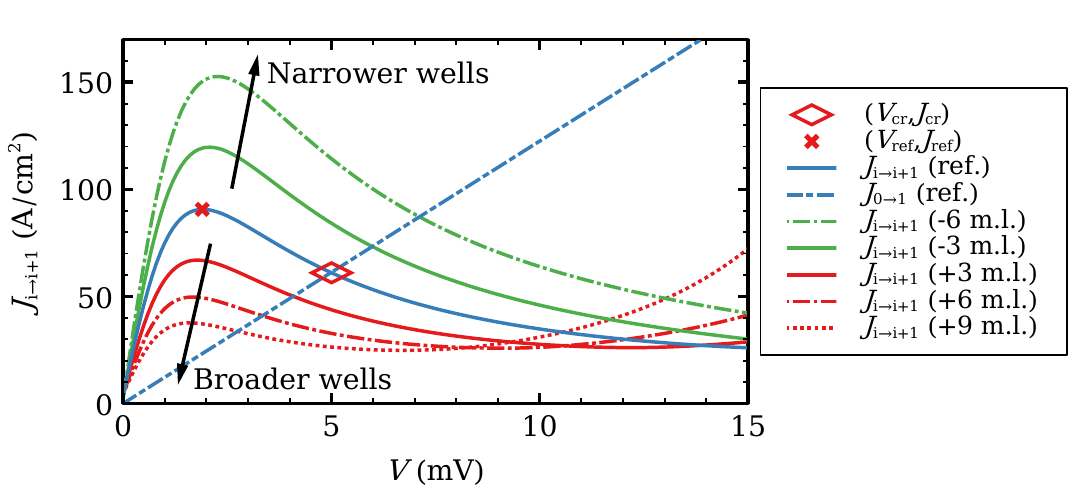}
\end{center}
\caption{Tunneling current-voltage characteristics for the ideal SSL with $n_i=N_D$, $V_i=V$ comparing the reference configuration (ref.) $d_B=4$ nm, $d_W=7$ nm to the contact Ohm's law (dot-dashed straight line) and to other configurations with more or less monolayers (m.l.) at the wells. The rhombus marks the critical current $J_\text{cr}$ and voltage$V_\text{cr}$ at which the contact Ohm's law intersects the reference configuration. Reprinted from E. Momp\'o, M. Carretero, L. L. Bonilla, Designing hyperchaos and intermittency in semiconductor superlattices, Physical Review Letters 127, 096601 (2021); https://doi.org/10.1103/PhysRevLett.127.096601} \label{fig3} 
\end{figure}

\subsection{Effect of one wider well on the $I-V$ characteristics}
What can we expect by modifying the width of a well in an otherwise ideal SSL? Let us consider the tunneling current of an ideal SSL for fixed electron densities $n_i=N_D$ and a constant barrier voltage drop $V$ in Eq~\eqref{eq18}, $J_{i\to i+1}(V)$. Fig.~\ref{fig3} shows the curve for the SSL reference configuration ($d_B=4$ nm, $d_W=7$ nm),  the curves when we add or subtract a number of monolayers (0.3 nm wide each) to $d_W$, and Ohm's law at the contact, $J_{0\to 1}(V)$. $J_{i\to i+1}(V)$ exhibits a single maximum at the shown voltage range. Widening the well decreases the maximum and shifts it toward lower voltages. The opposite occurs when we have narrower wells. The intersection of $J_{i\to i+1}(V)$ and $J_{0\to 1}(V)$ (marked with a rhombus for the reference configuration) changes accordingly. This intersection roughly marks the voltage and current at which the contact issues a dipole wave, which is the mechanism behind self-sustained oscillations of the current, excitability and other phenomena \cite{bon02r,wac02,bon05,BT10}.

If we have a long ideal SSL at the reference configuration, the stationary voltage profile for a fixed current lower than $J_\text{ref}$, will be a solution $V^{(1)}(J)$ of $J_{i\to i+1}(V)=J$ on the first branch of $J_{i\to i+1}(V)$, except for a short region near the emitter where $V_0$ decreases from $V_0=J/J_{0\to 1}$ to $V^{(1)}$. Dynamics of one charge dipole wave occurs as follows. When the wave is far from the contacts, the field profile is roughly constant outside the wave except near the emitter. As the wave arrives at the collector, the current increases with time so as to keep the voltage at its constant value. If $J(t)$ surpasses $J_\text{ref}$, the stationary state is no longer stable and a dipole wave is injected at the emitter contact. Repetition of this phenomenon produces the self-oscillations of the current \cite{bon02r,wac02,bon05,BT10}. Now, suppose we insert a different well far from the contacts in the reference configuration. The intersection of $J_{i\to i+1}(V)$ and $J_{0\to 1}(V)$ occurs at lower (resp. higher) current for a wider (resp. narrower) well than the reference one. Then we may expect that inserting a wider well may facilitate triggering a dipole wave in it when the current surpasses the corresponding intersecting value. The opposite is true if the inserted well is narrower. Thus, we expect richer SSL dynamics inserting wider wells.

We test our expectations by numerically simulating the deterministic model equations. With respect to the $I-V$ curve of the ideal SSL in Fig.~\ref{fig1}, each added monolayer shifts significantly the region of self-oscillations until there are 6 extra monolayers in total. From that point on, adding more monolayers to the modified well does not change the self-oscillation region of the $I-V$ curve. As in the case of unmodified SSLs, current self-oscillations are due to the dynamics of charge dipole waves. These waves are changed slightly when traveling through the wider well, which affects the evolution of the total current density by producing sudden and short-lived spikes.

\begin{figure}[htbp]
\begin{center}
\includegraphics[width=14cm,angle=0]{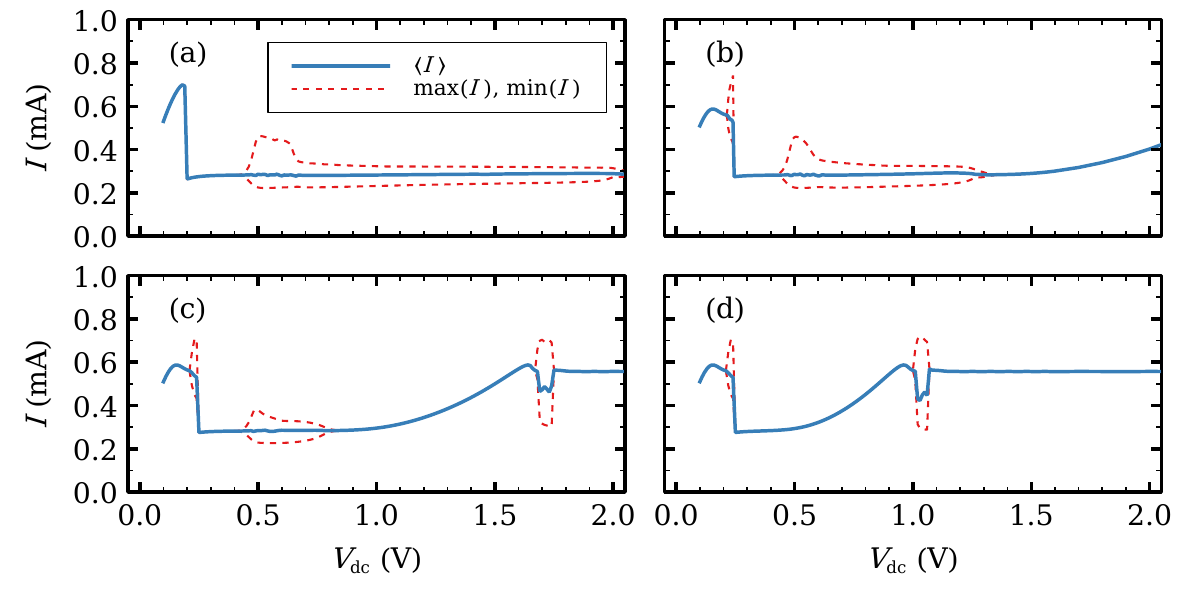}
\includegraphics[width=7cm,angle=0]{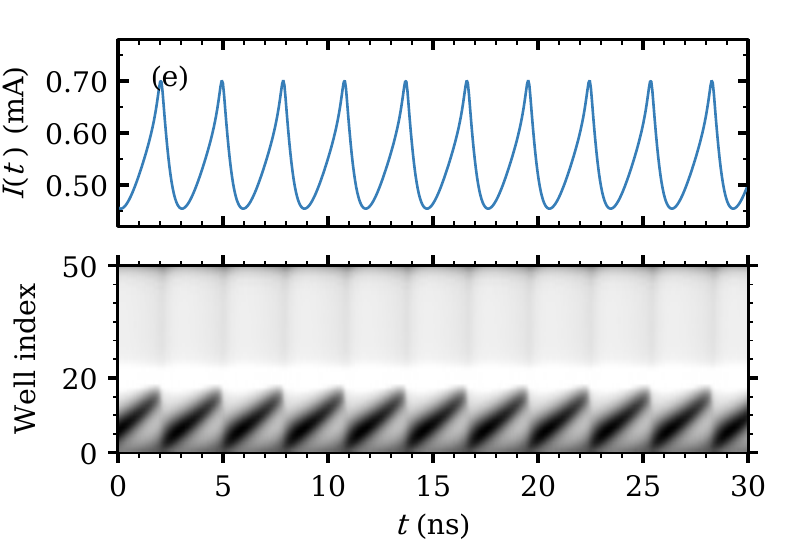}~~~
\includegraphics[width=7cm,angle=0]{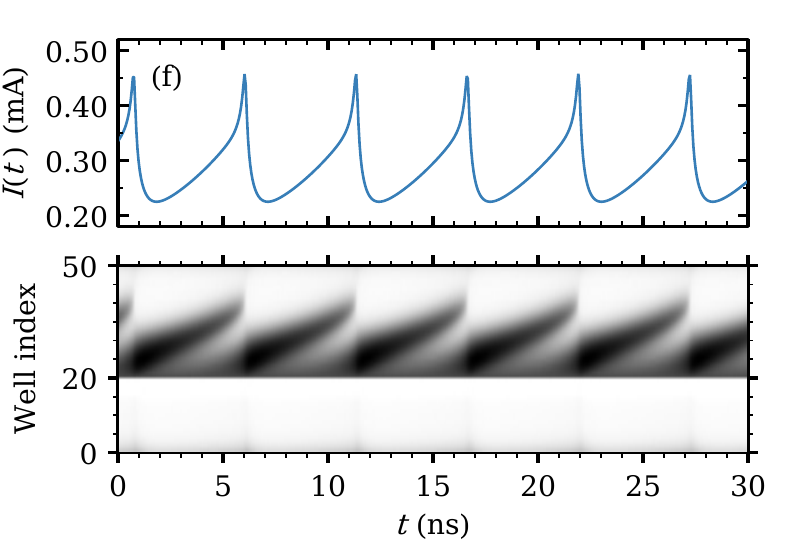}\\
\includegraphics[width=7cm,angle=0]{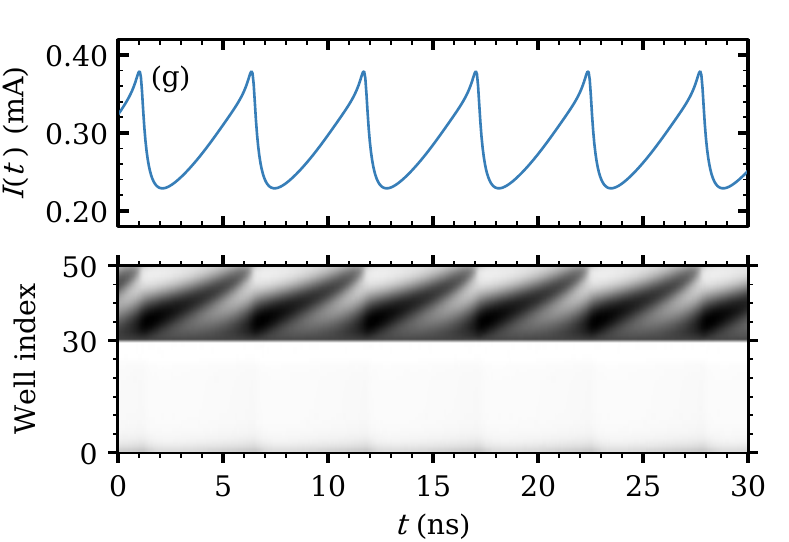}~~~
\includegraphics[width=7cm,angle=0]{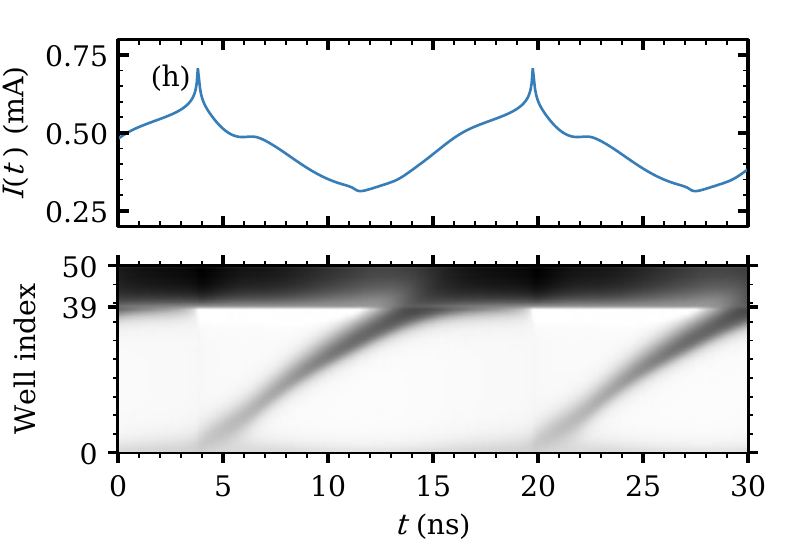}
\end{center}
\caption{First plateau of the current-voltage characteristics for the SSL with 50 identical periods when the modified 10 nm-wide well is at: (\textbf{a}) $i_w=5$, (\textbf{b}) $i_w=20$, (\textbf{c}) $i_w=30$, and (\textbf{d}) $i_w=39$. Current traces and density plots of the electric field for (\textbf{e}) \(i_w=20\), \(V_\text{dc}=0.23\,\text{V}\); (\textbf{f}) \(i_w=20\), \(V_\text{dc}=0.5\,\text{V}\); (\textbf{g}) \(i_w=30\), \(V_\text{dc}=0.5\,\text{V}\); (\textbf{h}) \(i_w=39\), \(V_\text{dc}=1.02\,\text{V}\). In the density plots, light and dark tones correspond to low and high field values, respectively.} \label{fig4} 
\end{figure}

We consider a SSL with a single modified well having 10 extra monolayers (i.e., $d_W= 10$ nm).  Its energy levels given by solving Eq.~\eqref{eq4} are $\mathcal{E}_{C_1}=24.0$ meV, $\mathcal{E}_{C_2}=96.1$ meV, $\mathcal{E}_{C_3} =214.7$ meV. The features of the $I-V$ curve depend on the location of the modified well and the general \emph{low voltage} behavior is the following. For voltages just above the onset of oscillations, dipole waves are repeatedly nucleated at the injector and disappear after a short trip. If a dipole wave born at the emitter can reach the wider well, no other wave will come out the emitter. Instead, self-sustained nucleations of dipole waves occur at the wider well for a large enough voltage bias $V_{dc}$. This fact plus the minimum number of SSL periods required for oscillations mean that the $I-V$ curve takes on different shapes depending on whether the wider well $i_w$ is near the injector, as in Fig.~\ref{fig4}(a), near the collector, as in Fig.~\ref{fig4}(d), or away from both contacts, as in Figs.~\ref{fig4}(b) and \ref{fig4}(c). 

To exhibit self-oscillations, SSLs need to surpass a critical length \cite{wac97} and their doping density should be smaller than a critical value \cite{wac97,car00}. In the parameter range explored in our numerical simulations, the minimum length for a traveling dipole wave to induce self-oscillations is 14 periods. For $i_w<14$, self-oscillations occur for a large voltage interval and are due to recycling at $i_w$, cf. Fig.~\ref{fig1}(a). For $i_w>14$ and as the dc voltage increases, the shape of the $I-V$ curve is as follows: 
\begin{enumerate}
	\item[(i)] There is a narrow voltage interval where charge waves nucleate at the emitter and die before reaching the modified well, cf. Figs.~\ref{fig4}(b) and \ref{fig4}(e).
	\item[(ii)] For larger $V_{dc}$, a dipole wave nucleated at the injector reaches $i_w$ and dies there. A stationary state forms for which excess charge is mostly located at $i_w$. This stationary state is accompanied by a large drop in the total current density, as shown in Figs.~\ref{fig4}(a) to \ref{fig4}(d). The current drop occurs because most charge accumulates at the emitter contact (high current state) for lower {\em dc} voltage but it accumulates at the modified QW (low current state) for larger {\em dc} voltage.
	\item[(iii)] If $i_w$ is between $i=14$ and $N-14$, current self-oscillations occur again at another voltage interval, as illustrated by Figs.~\ref{fig4}(b) and \ref{fig4}(c). This interval starts and ends in a supercritical Hopf bifurcation, and it becomes smaller as $i_w\to N-14$ and dipole waves nucleate at $i_w$. For larger voltages, the current is stationary and it rises smoothly to previous levels. As $i_w$ approaches $N-14$, self-oscillations occur for a third voltage range, depicted in Fig.~\ref{fig4}(c). In this narrow higher voltage interval, oscillations start as a supercritical Hopf bifurcation but end at a SNIPER bifurcation. In this high voltage range, dipole waves nucleate at the injector and travel towards $i_w$ whereas the electric field profile is large and quasi-stationary between $i_w$ and the collector; see Figs.~\ref{fig4}(f) and \ref{fig4}(g). After the SNIPER, the current becomes stationary and rises smoothly to second plateau levels.
	\item[(iv)] For $i_w$ between $N-14$ and $N$, the intermediate voltage range of self-oscillations disappears and only narrow voltage ranges corresponding to recycling of small waves traveling near the injector or the collector survive, cf. Figs.~\ref{fig4}(d) and \ref{fig4}(h).
\end{enumerate}

\begin{figure}[htbp]
\begin{center}
\includegraphics[width=14cm,angle=0]{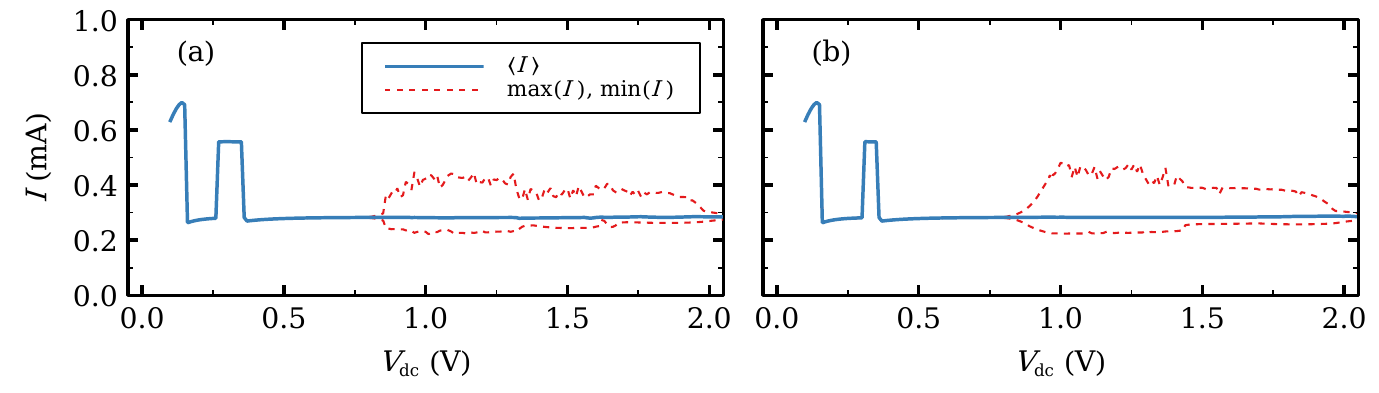}
\end{center}
\caption{First plateau of the current-voltage characteristics for the SSL with 50 identical periods with modified 10 nm-wide wells at $i_1=5$ and: (\textbf{a}) $i_2=25$, (\textbf{b}) $i_2=30$. The presence of two modified wells gives rise to two peaks in the low voltage, stationary, part of the $I-V$ characteristic curve. In case (a), self-oscillations are time periodic. In case (b), self-oscillations are time periodic for $V_{dc}<1$ V and for $V_{dc}>1.5$ V, and they are complex (mostly chaotic) for $1<V_{dc}<1.5$ V.} \label{fig5} 
\end{figure}

\subsection{Two wider wells}
Let us place two identical and wider wells at $i_1$ and $i_2$ ($i_1<i_2$). If the wider wells are different, the resulting dynamics will be similar to that explained previously for one well, for one of the modified wells will dominate. Let the widths of the wider wells be $d_{W_j}$, $j=1,2$, and let regions I, II and III be the intervals $i<i_1$, $i_1<i<i_2$, and $i>i_2$, respectively. As before, these wells have to include at least 6 extra monolayers for, otherwise, SSLs with less monolayers have drastically different $I-V$ curves. We shall fix $i_1=5$, so that dipole nucleation occurs at $i_1$ and not at the injector, as mentioned in relation to Fig.~\ref{fig4}(a). We then vary $i_2$. If $d_{W_1}> d_{W_2}$, charge dipoles nucleate at $i_1$ and travel through $i_2$ with a small disturbance. The situation is qualitatively similar to Fig.~\ref{fig4}(a) with $i_w=i_1=5$. If $d_{W_1}< d_{W_2}$, the situation is similar to that of a single wider well at $i_w=i_2$ with an injector at $i_1$. On the other hand, if $d_{W_1}\approx d_{W_2}$, SSL dynamics is more complex and interesting.

From now on, we consider $d_{W_1}= d_{W_2}=10$ nm. Fig.~\ref{fig5} illustrates typical $I-V$ characteristic curves. If regions II and III have more than 14 wells, dipole waves can be nucleated at $i_1$ and at $i_2$, they travel through regions II and III respectively, and their motion is strongly correlated. In general, each region II and III can support only one dipole wave. Some specific cases in which two waves may move on the same region will be discussed later. Correlation between dipole waves is as follows.
\begin{enumerate}
	\item[(i)] If a dipole wave does not reach the end of the region where it travels before it disappears, its annihilation will trigger nucleation at $i_1$ and at $i_2$. 
	\item[(ii)] Waves reaching $i_2$ (the end of region II) trigger nucleation at regions II and III.
	\item[(iii)] Waves reaching the end of region III will not necessarily trigger nucleation. These  waves can stop near the collector and stay there for either short or long times (and then recycle). This situation can be seen as a metastable state.
\end{enumerate}
Self-oscillations are time periodic in case (ii) but they may become chaotic in cases (i) and (iii), which explains the irregular disposition of the maxima in Fig.~\ref{fig5}. For $i_1=5$, observation of chaotic attractors requires the second well to satisfy $28< i_2\leq 35$. Note that modified SSLs exhibit self-oscillations with faster frequencies than in ideal SSLs because the dipole waves causing them travel on shorter regions of the device.

\begin{figure}[htbp]
\begin{center}
\includegraphics[width=12cm,angle=0]{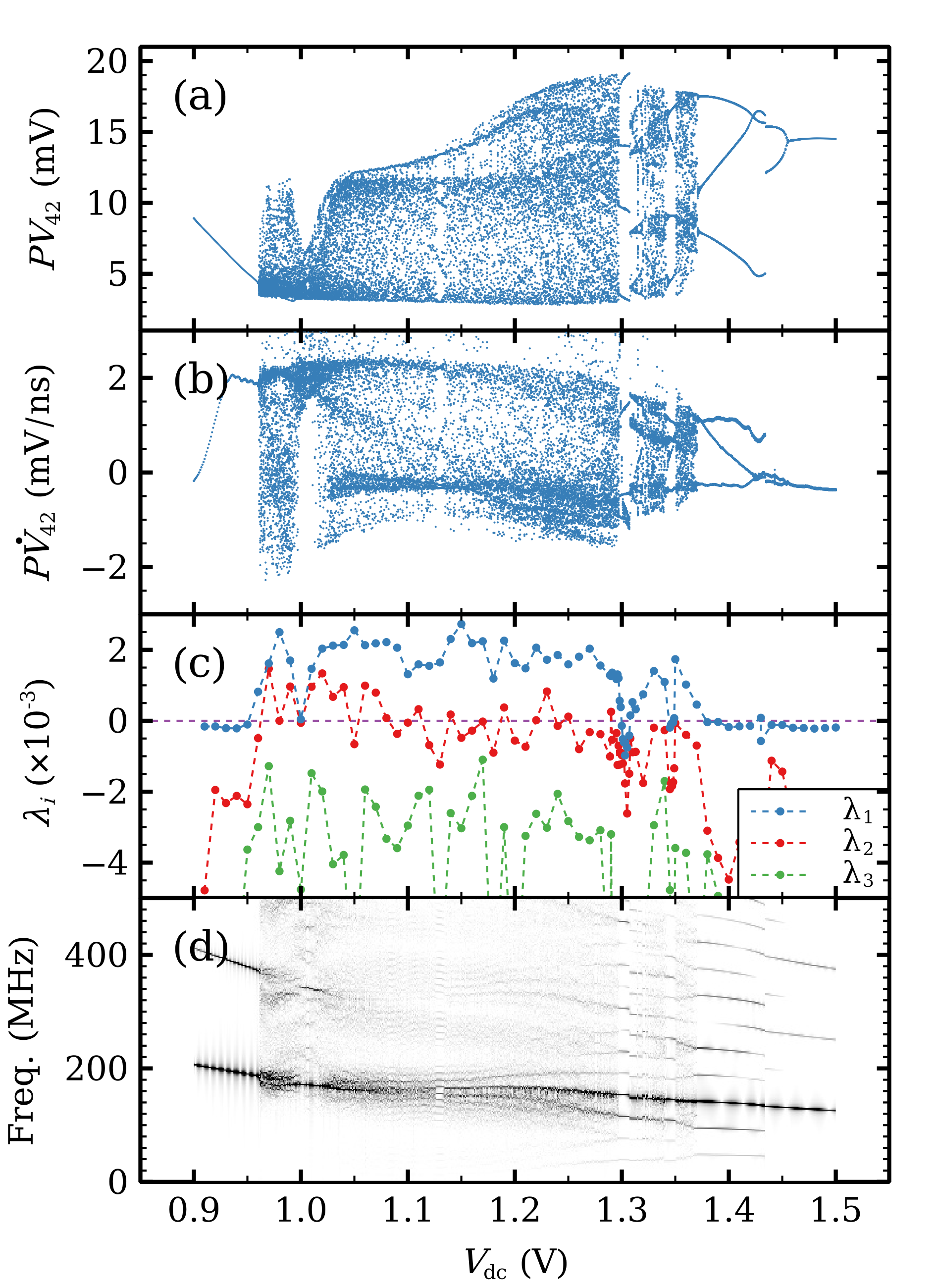}
\end{center}
\caption{(a) Poincar\'e map from $V_{42}(t)$, (b) Poincar\'e map from $\dot{V}_{42}(t)$, (c) Lyapunov exponents, and (d) Fourier spectrum as functions of dc voltage for the modified SSL with $i_1=5$ and $i_2=30$. Each panel shows features hidden in the other ones. The Poincar\'e map reveals jumps between periodic attractors at $V_{dc}=1.3\text{V}$ and $V_{dc}=1.43\text{V}$. The Fourier spectrum reveals the underlying behavior to be quasi-periodic with different incommensurate frequencies, whereas the Lyapunov exponents show that the system is hyperchaotic for $V_{dc}<1.08\text{V}$ ($\lambda_1, \lambda_2>0$ and of comparable scales). For $V_{dc}>1.08\text{V}$, the system has intermittent chaos at different time-scales ($\lambda_1\gg\lambda_2\approx 0$). Reprinted from E. Momp\'o, M. Carretero, L. L. Bonilla, Designing hyperchaos and intermittency in semiconductor superlattices, Physical Review Letters 127, 096601 (2021); https://doi.org/10.1103/PhysRevLett.127.096601} \label{fig6} 
\end{figure}

\section{Hyperchaos and intermittency} \label{sec:6}
In this section, we explore complex self-oscillations occurring in the SSL with modified wells of 10 nm width at $i_1=5$ and $i_2=30$, which has the $I-V$ curve depicted in Fig.~\ref{fig5}(b). 

Fig.~\ref{fig6} shows a variety of dynamical behaviors for the voltage range where self-oscillations occur in Fig.~\ref{fig5}(b). Each panel in Fig.~\ref{fig6} provides complementary information. The Poincar\'e maps in Figs.~\ref{fig6}(a) and \ref{fig6}(b) are constructed from the time traces of two well-separated periods, $V_{12}(t)$ and $V_{42}(t)$. Figs.~\ref{fig6}(a) and \ref{fig6}(b) depict the values of $V_{42}(t)$ and of $\dot{V}_{42}(t)$, respectively, at times $t^*$ where $V_{12}(t)$ takes on its mean value in time and $\dot{V}_{12}(t^*)>0$ (so as to avoid redundant symmetric points). Fig.~\ref{fig6}(c)  shows the three largest Lyapunov exponents. Fig.~\ref{fig6}(d) depicts the density plot of the normalized Fourier spectrum for each voltage value, which exhibits the dominant frequencies at each {\em dc} voltage. 

\begin{figure}[htbp]
\begin{center}
\includegraphics[width=12cm,angle=0]{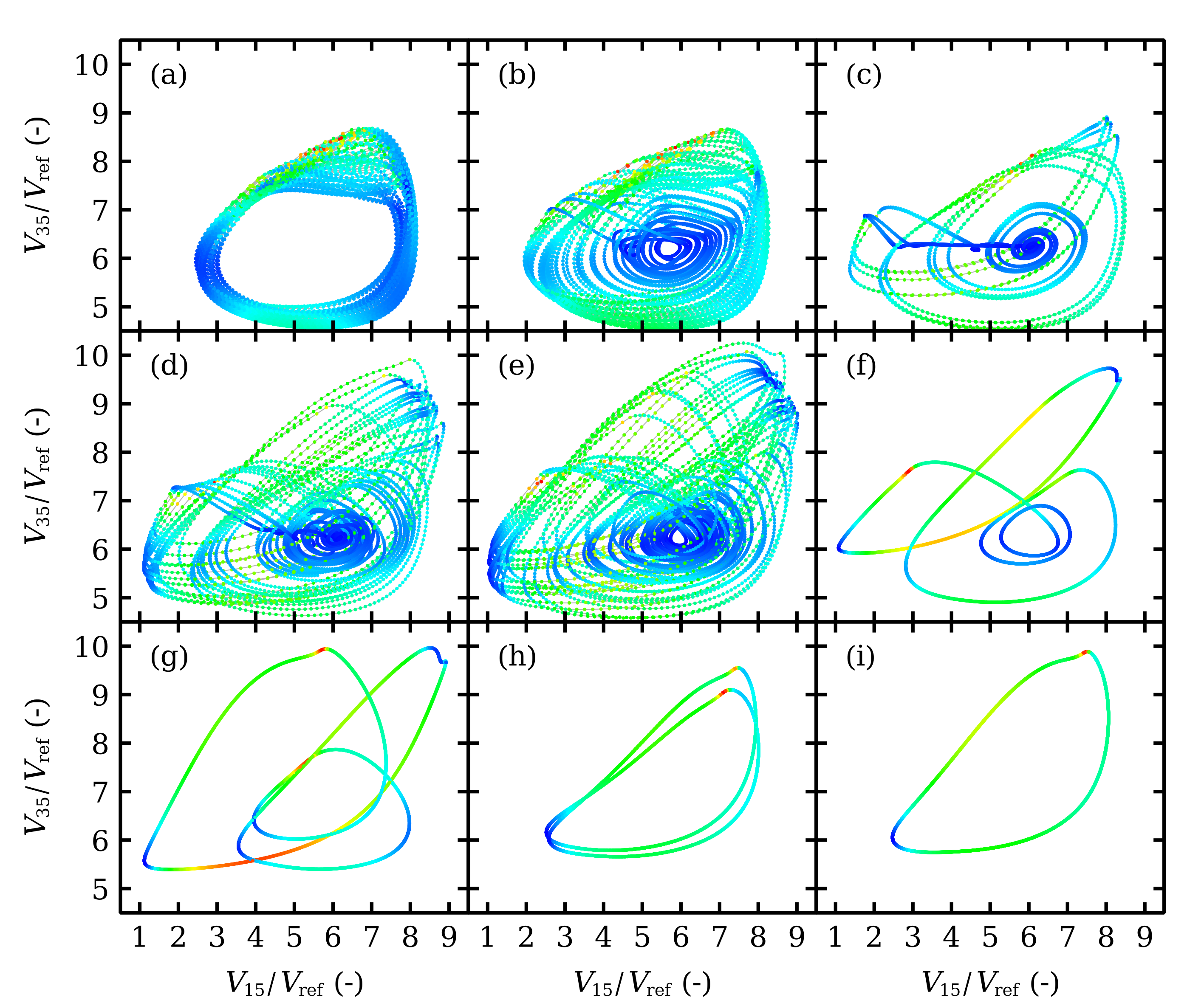}
\end{center}
\caption{Phase plane portraits $(V_{15},V_{35})$ for $V_{dc}=$ (\textbf{a}) 1.01 V, (\textbf{b}) 1.03 V, (\textbf{c}) 1.10 V, (\textbf{d}) 1.20 V, (\textbf{e}) 1.275 V, (\textbf{f}) 1.30 V, (\textbf{g}) 1.40 V, (\textbf{h}) 1.45 V, (\textbf{i}) 1.50 V. } \label{fig7} 
\end{figure}

\begin{figure}[htbp]
\begin{center}
\includegraphics[width=12cm,angle=0]{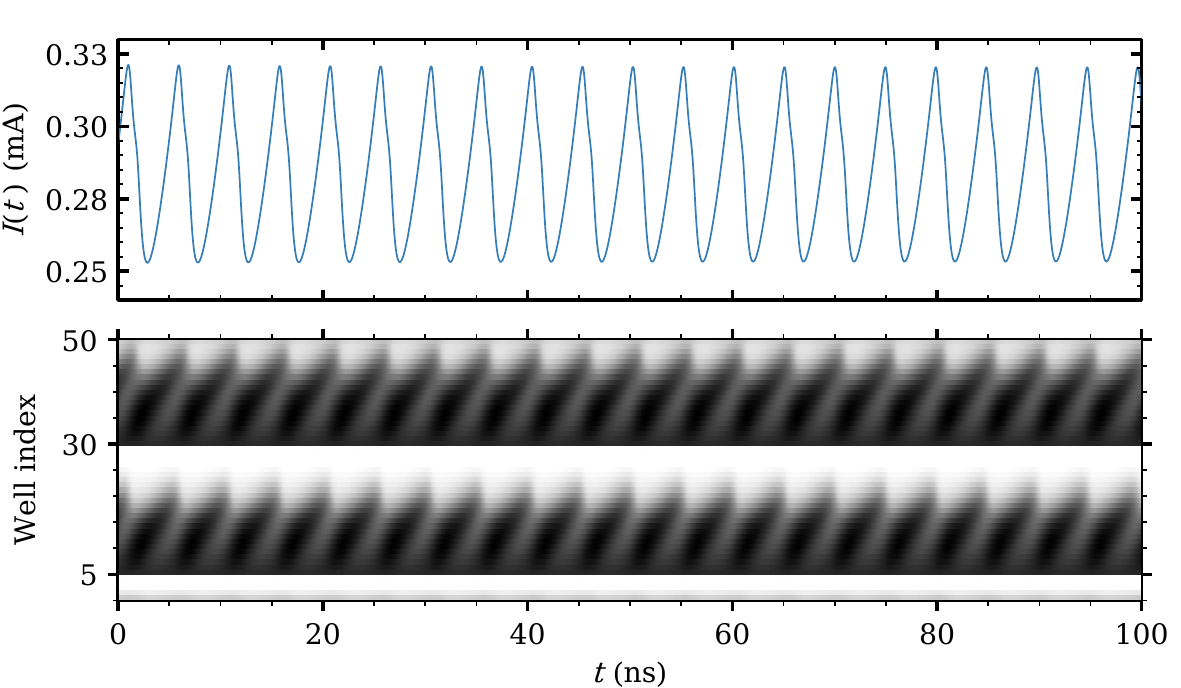}
\end{center}
\caption{Current traces and density plots of the electric field profile for $V_{dc}=0.9$ V. For this low voltage periodic attractor, the waves at regions II and III do not reach $i_2$ or the collector, respectively. In the density plots, light and dark tones correspond to low and high field values, respectively. Reprinted from E. Momp\'o, M. Carretero, L. L. Bonilla, Designing hyperchaos and intermittency in semiconductor superlattices, Physical Review Letters 127, 096601 (2021); https://doi.org/10.1103/PhysRevLett.127.096601} \label{fig8} 
\end{figure}

As $V_\text{dc}$ increases, the different attractors can be visualized by trajectories in the phase plane $(V_{15},V_{35})$ for the voltage drops in two widely separated barriers, cf Fig.~\ref{fig7}. Firstly, the stationary state loses its stability and a time periodic attractor appears at $V_\text{dc}=0.8$ V. The voltage profiles consist of charge dipole waves being repeatedly nucleated at both modified wells and advancing towards the collector without reaching it, cf Fig.~\ref{fig8}. 

\begin{figure}[htbp]
\begin{center}
\includegraphics[width=8cm,angle=0]{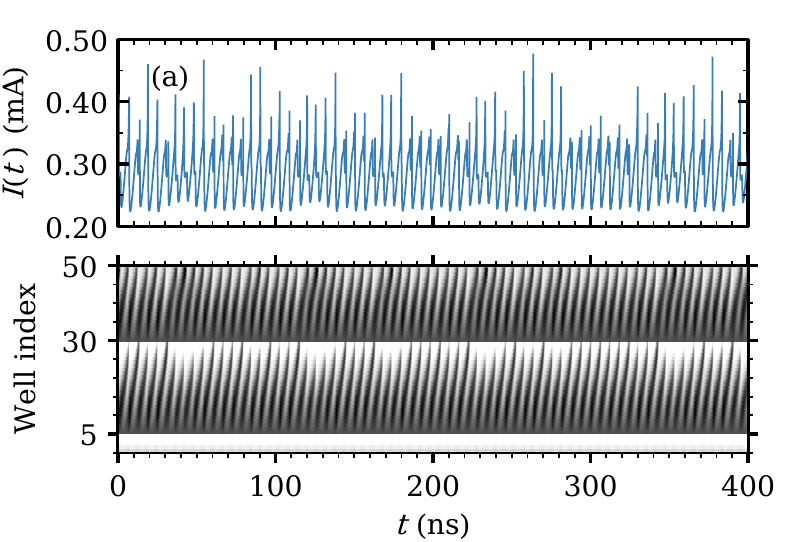}
\includegraphics[width=8cm,angle=0]{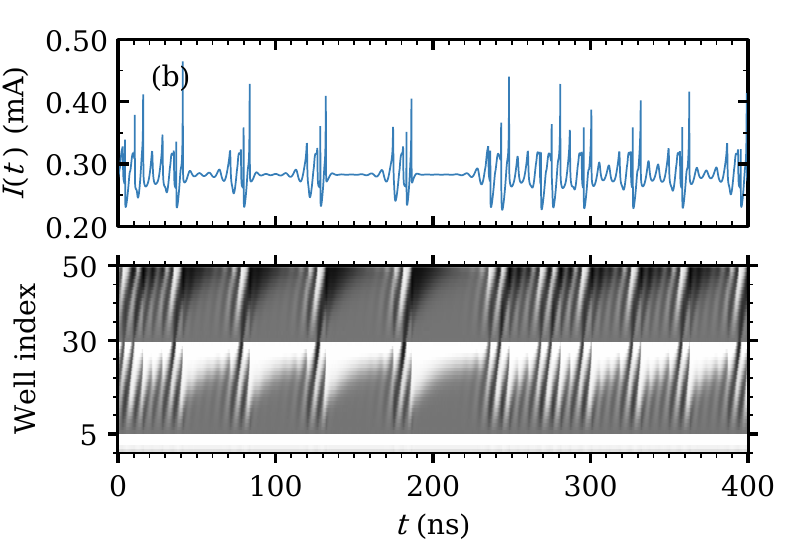}
\end{center}
\caption{Current traces and density plots of the electric field for (\textbf{a}) $V_{dc}=1.01$ V (hyperchaos, two positive Lyapunov exponents) and (\textbf{b}) $V_{dc}=1.20$ V (intermittency, one positive and one zero Lyapunov exponent). In the density plots, light and dark tones correspond to low and high field values, respectively. Reprinted from E. Momp\'o, M. Carretero, L. L. Bonilla, Designing hyperchaos and intermittency in semiconductor superlattices, Physical Review Letters 127, 096601 (2021); https://doi.org/10.1103/PhysRevLett.127.096601} \label{fig9} 
\end{figure}

At $V_\text{dc}=0.96$ V a second cycle (periodic attractor) appears and interacts with the first one. The result is a hyperchaotic attractor with two positive Lyapunov exponents. Trajectories fill the space between the two cycles. In the voltage interval $0.961 < V_\text{dc} < 1.1$ for hyperchaos (only one Lyapunov exponent is positive for $0.96<V_\text{dc}<0.961$), dipole waves nucleated at the second modified well either cannot reach the collector or, if they do, dipoles cannot stay in the wells near the collector. See Fig.~\ref{fig9}(a). For larger voltages, the second Lyapunov exponent becomes smaller albeit positive, and intermittent chaos appears instead, cf. Fig.~\ref{fig6}(c). This corresponds to the appearance of another cycle that interacts with the others and eventually disappears at a saddle point, as shown in Fig.~\ref{fig7} for $V_\text{dc}=1.03$ V and 1.10 V. Intermittency chaos appears for the interval $1.10 < V_\text{dc} < 1.37$: irregular bursts corresponding to a cycle are separated by intervals for which the trajectories are close to the saddle point, cf Fig.~\ref{fig7}. This behavior is associated to dipole waves that reach the collector, stop there and remain in the last SSL periods (quiescent stage), whereas periodic bursts are associated to dipole wave recycling in Regions II and III. At $V_\text{dc}=1.2$ V, the saddle point expands to a saddle cycle and the intermittent behavior continues. The difference is that the quiescent stage is associated to low frequency oscillations. See Fig.~\ref{fig9}(b). 

At $V_\text{dc}=1.37$ V, the intermittency becomes a period 3 cycle (three loop trajectories in the phase plane). At larger {\em dc} voltages the periodic behavior continues and it becomes simpler (two loops at 1.43 V, a single loop for larger voltages). The transition from periodic attractors with three loops to two loop ones at 1.43 V is rather abrupt, as shown in Figs.~\ref{fig6}(a) and \ref{fig6}(b). We have checked that there is a hysteresis cycle about this voltage value that becomes manifest by sweeping up or down the {\em dc} voltage. The last branch of time periodic oscillations disappears at a supercritical Hopf bifurcation.

\section{Effect of width randomness and noise}\label{sec:7}
\subsection{Width randomness}
When growing SSLs, it is difficult to control perfectly the width of the layers of the two semiconductors. In this section, we report the effects of fluctuations in well width on SSL current self-oscillations. We consider that the largest effects come from errors in the number of monolayers during epitaxial growth of the SSL. Thus, we ignore smaller effects such as fluctuations in barrier width and composition, internal and external noise and fluctuations in doping density. We set $d_{W_i}+\delta_i$ to be well widths, where $d_{W_5}=d_{W_{30}}=10$ nm, and $d_{W_i}=7$ nm for the other wells, $i\neq 5,30$. We extract $\delta_i$ out of a zero mean normal distribution with standard deviation $\sigma$. Then deviations larger than $\pm 2\sigma$ are rare. For example, if $\sigma=0.5$ nm, widths that deviate more than 1 nm from $d_{W_i}$ are rare. 

\begin{figure}[htbp]
\begin{center}
\includegraphics[width=8cm,angle=0]{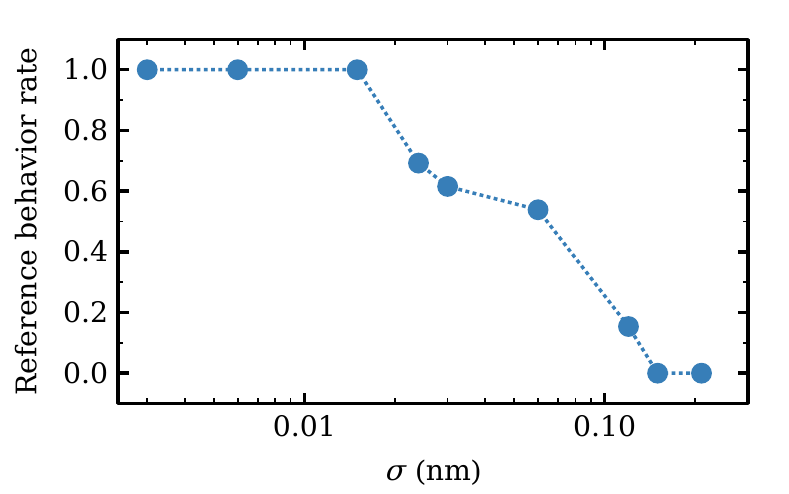}
\end{center}
\caption{Success rate measuring fraction of simulations where chaotic attractor remains for a given value of the standard deviation $\sigma$. Reprinted from E. Momp\'o, M. Carretero, L. L. Bonilla, Designing hyperchaos and intermittency in semiconductor superlattices, Physical Review Letters 127, 096601 (2021); https://doi.org/10.1103/PhysRevLett.127.096601} \label{fig10} 
\end{figure}

Given a random configuration of $\delta_i$ with standard deviation $\sigma$, we have numerically solved the SSL model. Depending on the obtained configuration, we have observed that intervals of hyperchaos or intermittent chaos are destroyed or still remain for that disordered configuration. When there are long voltage intervals (having widths comparable to those in the SSL without disorder, e.g., almost 1 V wide, as seen in Fig.~\ref{fig5}(b)) where the chaotic behavior of the SSL without disorder is kept undisturbed, we consider these examples as successes. If disorder causes new periodicity or stationary windows to appear within long voltage intervals of formerly chaotic behavior of the SSL without disorder, we consider these examples as failures. For a given value of $\sigma$, the success rate of disordered SSLs that still exhibit chaotic behavior is shown in Fig.~\ref{fig10}. For $\sigma<0.015$ nm, chaotic attractors observed for the SSL without disorder remain. However, $\sigma= 0.024$ nm is sufficient to have a lower success rate of 70\%. 

A different observation is the following. Let $(x_1, \ldots,x_N)$ be a sequence of numbers obtained from a normal distribution with zero mean and unit variance. Then $\delta_i=\sigma x_i$ correspond to a configuration extracted from a normal distribution with zero mean and $\sigma^2$ variance. By increasing $\sigma$, we find a threshold $\sigma_\text{cr}(x_1, \ldots ,x_N)$, above which the attractors of the SSL without disorder undergo significant changes (e.g., inhibition or disappearance of current oscillations). This is related to the change of current-voltage characteristics with well width displayed in Fig.~\ref{fig4}. If the difference between widths of adjacent wells, $\delta_i-\delta_{i+1}>0$, is large, dipole waves experience difficulty crossing these wells. In turn, this explains why disorder inhibits oscillations and chaos.

When building SSL devices, disorder effects are very important and have to be controlled as much as possible. During epitaxial growth \cite{gra95}, Al atoms within each interface alloy monolayer may be segregated into local clusters or not be positioned randomly in the Ga or the As sublattice \cite{das12}. This yields a nonzero $\sigma$ even if there are no errors in the number of monolayers per barrier and well (recall that the monolayer width is 0.3 nm). Careful design achieves $\sigma<0.018$ nm in simpler devices \cite{das12,mis15}, which would yield reliably chaotic SSLs according to the success rate of Fig.~\ref{fig10}.

\subsection{Noise}
\begin{figure}[htbp]
	\begin{center}
		\includegraphics[width=10cm,angle=0]{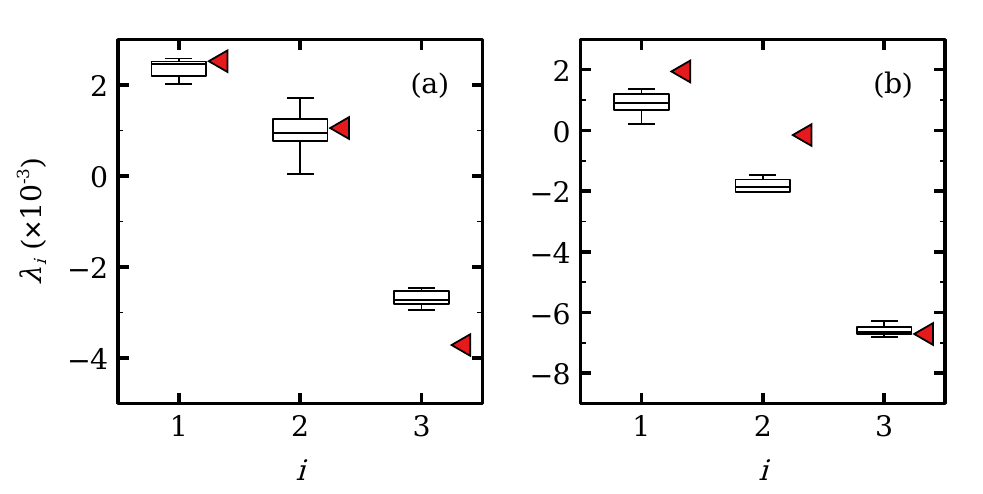}
	\end{center}
	\caption{Effect of shot and thermal noise on the three largest Lyapunov exponents (noiseless values marked by tilted triangles) for (\textbf{a}) \(V_{\text{dc}}=0.98\,\text{V}\) (hyperchaos) and (\textbf{b}) \(V_{\text{dc}}=1.15\,\text{V}\) (intermittent chaos). The boxes describe the distribution of exponents and the vertical bars indicate the standard deviation of fluctuations.} \label{fig11} 
\end{figure}

Here we consider the effects of internal noise (shot and thermal noise), as in Eqs.~\eqref{eq21} and \eqref{eq23}, and the effect of external voltage noise (2 mV rms for a 50 Ohm resistor), as in Eq.~\eqref{eq22}, on the Lyapunov exponents of chaotic attractors. Fig.~\ref{fig11} shows our results. For hyperchaos, noise produces a dispersion near the deterministic values of the Lyapunov exponents, with larger standard deviation for the second largest exponent, cf Fig.~\ref{fig11}(a). For intermittent chaos, the two largest Lyapunov exponents are noticeably smaller than their values in the absence of noise cf Fig.~\ref{fig11}(b). Thus, by repeated simulations, we have concluded that the effect of noise is to decrease the largest Lyapunov exponent of the chaotic attractors and to increase slightly the third Lyapunov exponent. The latter does not become positive. Noise forces the system to visit more often contraction regions of the phase space such as the quiescent regions between bursts in intermittent chaos. This lowers the largest Lyapunov exponent \cite{zho02}. Thus, contrary to the effect reported and observed in ideal SSL with identical periods \cite{alv14,yin17}, noise does not enhance chaos in these modified SSL, but its effect is quite small.

\section{Conclusions}\label{sec:8}
In this paper, we have presented a general theory of nonlinear vertical transport in weakly coupled SSLs based on the large separation of time scales \cite{bon94}: the relaxation times within subbands are much shorter than intersubband scattering times which, in turn, are much shorter than dielectric relaxation times. From this hierarchy of times, we derive spatially discrete equations for voltage drops at barriers and wells and subband populations. When intersubband scattering times are much shorter than dielectric relaxation of electrons, we obtain the sequential resonant tunneling model of Ref.~\cite{agu97} which can be further simplified to equations for the average electric fields and electron densities at QWs \cite{bon00,bon02r}. Other theories yield different expressions for the tunneling current \cite{wac02} but they still assume the same hierarchy of time scales.

The $I-V$ current-voltage characteristics of the SSL provides an equivalent bifurcation diagram of stable solutions of its governing equations. It is extremely sensitive to the chosen configuration and to the fluctuations in well widths due to epitaxial growth of the SSL \cite{gra95}. Our sequential resonant tunneling equations have been modified to take into account these fluctuations as well as internal and external noises. We start considering an ideal SSL of identical periods without noise effects. Typically, it has stationary states (whose field profile consists of a low and a high field domain joined by some intermediate fields which constitute a frozen wavefront) and time periodic states due to the motion of high field domains that are traveling charge dipoles. These dipole waves are repeatedly triggered at the emitter contact and move toward the collector contact of the SSL \cite{bon02r,bon05,BT10}. To produce a richer dynamics, we have inserted one wider QW in the SSL because this allows for triggering additional dipole waves at the modified well depending on its location and applied {\em dc} voltage bias. Indeed, we have found richer $I-V$ characteristics and larger voltage intervals of time-periodic solutions, but we have not found chaotic solutions. Inserting two wider wells of different widths will result in dynamics similar to the one wider well case because the wider of the two wells dominates dynamics. Inserting two identical wider wells produces a more robust and resilient chaos on wider bias ranges: hyperchaos with two positive Lyapunov exponents and intermittent chaos with a single positive exponent \cite{mom21}.

Chaotic states of deterministic dynamics for two identically modified wells persist if we add realistic internal and external noises and are robust to sufficiently small disorder fluctuations. If the difference between widths of adjacent wells due to disorder is large, dipole waves experience difficulty crossing these wells. This impediment for dipole waves to move across the SSL explains why disorder inhibits oscillations and chaos. Thus, there is a competition between chaotic dynamics of the deterministic equations requiring dipole waves in two identically modified wells and disorder due to epitaxial growth that can localize charge dipoles at wells and forestall current oscillations. State of the art epitaxial growth techniques are known to produce devices with no errors in the number of monolayers per barrier and well and standard deviations smaller than 0.018 nm. According to Fig.~\ref{fig10}, such successful growth would produce reliably chaotic SSLs with a success rate over 70\%. It is plausible that inserting more identical wider wells on longer SSLs may increase the complexity of the resulting dynamics but there will be compromise between the total number of admissible SSL periods that is possible to grow with sufficiently small standard deviation and the errors introduced during epitaxial growth. The complex dynamics described in this paper could be observed in experiments with epitaxially grown SSLs. Another application is using synchronization of chaotic devices for secure communications \cite{kin10}. Since synchronization of chaotic SSLs has been demonstrated in experiments \cite{li15,liu18}, it is possible to use our work to build devices that distribute encryption keys safely by exploiting chaos synchronization \cite{keu17}.

\vspace{6pt} 


\acknowledgments
This work was funded by the FEDER/Ministerio de Ciencia, Innovaci\'on y Universidades -- Agencia Estatal de Investigaci\'on grant PID2020-112796RB-C22, by the Madrid Government (Comunidad de Madrid-Spain) under the Multiannual Agreement with UC3M in the line of Excellence of University Professors (EPUC3M23), and in the context of the V PRICIT (Regional Programme of Research and Technological Innovation). 
 





\appendix

\section{Derivation of deterministic tunneling currents}\label{app1}
Following Ref.~\cite{xu07}, the tunneling Hamiltonian is
\begin{eqnarray}
&&H_{\rm total}= H+H_T=\sum_{i=0}^{N+1}H_i + \sum_{j=0}^NH_{T_j},\label{a1}\\
&&H_i=\sum_{\mathbf{k}_i} E_{i\mathbf{k}_i} c^\dagger_{i\mathbf{k}_i} c_{i\mathbf{k}_i}, \quad 
H_{T_j}=\sum_{\mathbf{k}_j\mathbf{k}_{j+1}}\left( T_{\mathbf{k}_j\mathbf{k}_{j+1}} c_{j+1\mathbf{k}_{j+1}}^\dagger c_{j\mathbf{k}_j}+ H.c.\right)\!. \label{a2}
\end{eqnarray}
Here the Hamiltonian $H$ is a sum of individual Hamiltonians for each QW or contact and assumes that they are uncoupled from one another. H.c. stands for the Hermitian conjugate of the preceding term. The unperturbed single-electron states have absolute energies denoted by $E_{i\mathbf{k}_i}$ measured from the conduction band edge in the emitter contact. We have $E_{j\mathbf{k}_j}=\epsilon+E_\perp$, $E_\perp= \hbar^2\mathbf{k}_\perp^2/(2m_W)$, in which $\epsilon$ is the energy at the well, and $\mathbf{k}_\perp$ comprises the components of the wave vector that are orthogonal to the SL growth direction. The operators $c^\dagger_{i\mathbf{k}_i}$ and $c_{i\mathbf{k}_i}$ denote creation and annihilation operators for electrons in the $i$th well or contact with three-dimensional wave vector $\mathbf{k}_i$ and satisfy standard fermionic commutation rules: $\{c_{i\mathbf{k}_i}, c_{j\mathbf{k}_j}\}=c_{i\mathbf{k}_i} c_{j\mathbf{k}_j}+ c_{j\mathbf{k}_j}, c_{i\mathbf{k}_i}=0$, $\{c_{i\mathbf{k}_i}^\dagger, c_{j\mathbf{k}_j}^\dagger\}=0$, $\{c_{i\mathbf{k}_i}, c_{j\mathbf{k}_j}^\dagger\}=\delta_{ij}\delta_{\mathbf{k}_i\mathbf{k}_j}$. Each QW contains a set of $n$ subbands whose Fermi energy measured from the conduction band edge in the emitter contact are $\epsilon_{w_i^{(\nu)}}$, $\nu=1,\ldots,n$. $H_T$ is a small perturbation of $H$ representing the tunneling coupling between adjacent wells. Typically, an electron tunnels from the first subband of a QW into subband $\nu$ of the next QW and the electron population in the subbands is in local equilibrium because the relaxation times within subbands are much shorter than intersubband scattering times which, in turn, are much shorter than dielectric relaxation times. This hierarchy of times \cite{bon94} is the basis for all theories of spatially discrete models of SSLs \cite{bon05,bon02r,wac02,BT10}.

The change of the electron operator number at the $i$th well, $N_i=\sum_{\mathbf{k}_i} c^\dagger_{i\mathbf{k}_i} c_{i\mathbf{k}_i}$, is related to the tunneling current operator $\hat{J}_{i\to i+1}$ by
\begin{eqnarray}
e\dot{N}_i=  \frac{i}{\hbar}[H_{\rm total},eN_i] =\frac{i}{\hbar}[H_{T_{i-1}},eN_i]-\frac{i}{\hbar}[H_{T_{i}},eN_i] = \hat{J}_{i-1\to i}-\hat{J}_{i\to i+1}. \label{a3}
\end{eqnarray}
In the interaction representation, we have $H_T(t)=e^{iHt/\hbar} H_Te^{-iHt/\hbar}$ and $\hat{J}_{i\to i+1}(t)=e^{iHt/\hbar} \hat{J}_{i\to i+1}e^{-iHt/\hbar}$ and the average tunneling current density from subband 1 of QW $i$ to subband $\nu$ of QW $i+1$ satisfies the Kubo formula \cite{kub57,kub91,zwanzig}
\begin{eqnarray}
J_{1,i\to\nu, i+1}(t)= \frac{1}{\hbar}\int_{-\infty}^t \langle [\hat{J}_{1,i\to\nu, i+1}(t),H_T(t')] \rangle\, dt'. \label{a4}
\end{eqnarray}
Here the average is over the thermodynamic local equilibria at the subbands of QWs $i$ and $i+1$. A straightforward lengthy evaluation yields
\begin{eqnarray}
J_{1,i\to\nu, i+1}(t)&=& \frac{4\pi e}{\hbar}\sum_{\mathbf{k}_j\mathbf{k}_{j+1}} |T_{\mathbf{k}_j\mathbf{k}_{j+1}}|^2 \delta(E_{j+1\mathbf{k}_{j+1}}-E_{j\mathbf{k}_j})\nonumber\\
&\times& [n_F(E_{j+1\mathbf{k}_{j+1}}-\epsilon_{w_{j+1}})-n_F(E_{j\mathbf{k}_j}-\epsilon_{w_j})], \label{a5}
\end{eqnarray}
where $n_F(x)=1/(1+e^{x/k_BT})$ is the Fermi distribution function. The matrix element:
\begin{eqnarray}
T_{\mathbf{k}_j\mathbf{k}_{j+1}}=\frac{\hbar^2}{2m_B}\int_A (\psi_j\nabla\psi_{j+1}^*-\psi^*_{j+1}\nabla\psi_{j})\cdot d\mathbf{A}, \label{a6}
\end{eqnarray}
is calculated by using Bardeen's Transfer Hamiltonian method \cite{bardeen,pay86}. The wave functions of two adjacent square QWs, $\psi_j$ and $\psi_{j+1}$, are approximated by those of free particles in two isolated wells separated by an infinitely thick barrier. Then continuity of wave functions and their derivatives are used to find out the coefficients of the wave function expressions in different space intervals and the resulting wave functions produce the matrix element \eqref{a6} \cite{pay86}. The result is \cite{xu07}
\begin{eqnarray}
|T_{\mathbf{k}_i\mathbf{k}_{i+1}}|^2=\frac{\pi\hbar^4}{2m_B^2} B_{i-1,i}B_{i,i+1}T_i \delta_{\mathbf{k}_\perp\mathbf{k}_\perp '}, \label{a7}
\end{eqnarray}
where the coefficients $B_{i-1,i}$ and the transmission coefficient are given by Eqs.~\eqref{eq7} and \eqref{eq8}, respectively. We now transform the sums in \eqref{a5} to integrals over the energies $E_{j\mathbf{k}_j}=\epsilon+ E_\perp$ using a broadened spectral density to account for scattering. If the latter depends only on $\epsilon$, we obtain
\begin{eqnarray}
&&J_{1,i\to\nu, i+1}= \frac{e\hbar}{2m_B} \int d\epsilon\, A_{C1}^i(\epsilon)A_{C\nu}^{i+1}(\epsilon)B_{i-1,i}(\epsilon)B_{i,i+1}(\epsilon) T_i (\epsilon)\nonumber\\
&&\quad\quad\quad\times\!\int\! dE_\perp\!\! \left[\frac{1}{1+e^{(\epsilon+E_\perp-\epsilon_{w_i}})/k_BT}+\frac{1}{1+e^{(\epsilon+E_\perp-\epsilon_{w_{i+1}})/k_BT}} \right]\!\!. \label{a12}
\end{eqnarray}
After changing variables so that the energy is measured from the bottom of the QW $i$ \cite{bon02r}, we perform the integral over $E_\perp$, and get Eq.~\eqref{eq6} as a result.



\end{document}